\documentclass[pdftex,twocolumn,epjc3]{svjour3}          
\RequirePackage[T1]{fontenc}
\smartqed  
\RequirePackage{graphicx}
\RequirePackage{flushend}
\RequirePackage[numbers,sort&compress]{natbib}
\RequirePackage[colorlinks,citecolor=blue,urlcolor=blue,linkcolor=blue]{hyperref}

\journalname{}

\usepackage{cuted} 
\usepackage{graphicx,multirow,amsmath,amssymb,amsfonts,amsthm,mathrsfs,xcolor,textcomp,manyfoot,booktabs,algorithm,algorithmicx,algpseudocode,listings}
\usepackage{placeins} 
\begin{document}
{
\allowdisplaybreaks

\ifdefined\r \let\R\r \renewcommand{\r}[1]{(\ref{#1})}
\else                   \newcommand{\r}[1]{(\ref{#1})}
\fi
\ifdefined\tn   \renewcommand{\tn}[1]{\textnormal{#1}}
\else             \newcommand{\tn}[1]{\textnormal{#1}}
\fi
\ifdefined\rect \renewcommand{\rect}[1]{{\textstyle \frac{1}{#1}}}
\else             \newcommand{\rect}[1]{{\textstyle \frac{1}{#1}}}
\fi
\ifdefined\fract \renewcommand{\fract}[2]{{\textstyle \frac{#1}{#2}}}
\else              \newcommand{\fract}[2]{{\textstyle \frac{#1}{#2}}}
\fi
\ifdefined\soint \renewcommand{\soint}[4]{\hspace{#1}\oint_{#3}^{#4}\hspace{#2}}
\else              \newcommand{\soint}[4]{\hspace{#1}\oint_{#3}^{#4}\hspace{#2}}
\fi
\ifdefined\sointl \renewcommand{\sointl}[4]{\hspace{#1}\oint\limits_{#3}^{#4}\hspace{#2}}
\else               \newcommand{\sointl}[4]{\hspace{#1}\oint\limits_{#3}^{#4}\hspace{#2}}
\fi
\ifdefined\shointl \renewcommand{\shointl}[3]{\hspace{#1}\oint\limits^{#3}\hspace{#2}}
\else                \newcommand{\shointl}[3]{\hspace{#1}\oint\limits^{#3}\hspace{#2}}
\fi
\ifdefined\fracd \renewcommand{\fracd}[2]{\frac{\displaystyle{#1}}{\displaystyle{#2}}}
\else              \newcommand{\fracd}[2]{\frac{\displaystyle{#1}}{\displaystyle{#2}}}
\fi
\ifdefined\bs \renewcommand{\bs}[1]{\boldsymbol{#1}}
\else           \newcommand{\bs}[1]{\boldsymbol{#1}}
\fi
\ifdefined\td \renewcommand{\td}[2]{\frac{d{#1}}{d{#2}}}
\else           \newcommand{\td}[2]{\frac{d{#1}}{d{#2}}}
\fi  
\ifdefined\Eq \renewcommand{\Eq}[1]{Eq.~(\ref{#1})}
\else           \newcommand{\Eq}[1]{Eq.~(\ref{#1})}
\fi
\ifdefined\Eqs \renewcommand{\Eqs}[2]{Eqs.~(\ref{#1}) and (\ref{#2})}
\else            \newcommand{\Eqs}[2]{Eqs.~(\ref{#1}) and (\ref{#2})}
\fi
\ifdefined\sint   \renewcommand{\sint}[4]{\hspace{#1}\int_{#3}^{#4}\hspace{#2}}
\else               \newcommand{\sint}[4]{\hspace{#1}\int_{#3}^{#4}\hspace{#2}}
\fi
\ifdefined\slint  \renewcommand{\slint}[3]{\hspace{#1}\int_{#3}\hspace{#2}}
\else               \newcommand{\slint}[3]{\hspace{#1}\int_{#3}\hspace{#2}}
\fi
\ifdefined\slintl \renewcommand{\slintl}[3]{\hspace{#1}\int\limits_{#3}\hspace{#2}}
\else               \newcommand{\slintl}[3]{\hspace{#1}\int\limits_{#3}\hspace{#2}}
\fi
\ifdefined\sintl \renewcommand{\sintl}[4]{\hspace{#1}\int\limits_{#3}^{#4}\hspace{#2}}
\else              \newcommand{\sintl}[4]{\hspace{#1}\int\limits_{#3}^{#4}\hspace{#2}}
\fi
\ifdefined\rec   \renewcommand{\rec}[1]{\frac{1}{#1}}
\else              \newcommand{\rec}[1]{\frac{1}{#1}}
\fi
\ifdefined\z   \renewcommand{\z}[1]{\left({#1}\right)}
\else            \newcommand{\z}[1]{\left({#1}\right)}
\fi
\ifdefined\m   \renewcommand{\B}[1]{\mathbb{#1}}
\else            \newcommand{\B}[1]{\mathbb{#1}}
\fi
\ifdefined\m   \renewcommand{\m}[1]{\mathrm{#1}}
\else            \newcommand{\m}[1]{\mathrm{#1}}
\fi
\ifdefined\v   \renewcommand{\v}[1]{\mathbf{#1}}
\else            \newcommand{\v}[1]{\mathbf{#1}}
\fi
\ifdefined\c   \renewcommand{\c}[1]{\mathcal{#1}}
\else            \newcommand{\c}[1]{\mathcal{#1}}
\fi
\ifdefined\biggg \renewcommand{\biggg}[1]{\scalebox{1.2}{\Bigg{#1}}}
\else              \newcommand{\biggg}[1]{\scalebox{1.2}{\Bigg{#1}}}
\fi

\makeatletter

\title{A self-consistent calculation of non-spherical Bose--Einstein correlation functions with Coulomb final-state interaction}


\author{Márton I. Nagy\thanksref{addr1}
        \and
        Máté Csanád\thanksref{addr1}
        \and
        Dániel Kincses\thanksref{e3,addr1}
}

\thankstext{e3}{e-mail: kincses@ttk.elte.hu}

\institute{ELTE Eötvös Loránd University, Pázmány Péter sétány 1/A, Budapest, H-1117, Hungary\label{addr1}
}

\date{}

\maketitle

\begin{abstract}

Particle correlations and femtoscopy are a rich subfield of high-energy physics. As the experimental data become more precise, there is an increasing need for the theoretical calculations to provide better and more general descriptions of the measurements. One of the important new directions is the investigation of the precise shape of the Bose-Einstein correlation functions utilizing L\'evy-stable distributions. This work is a direct follow-up to our previous study, in which we developed a novel method for calculating Bose–Einstein correlation functions including the Coulomb final-state interaction. In this paper, we present a self-consistent generalization of the previous approach to non-spherical source functions and investigate the validity of the previously applied approximations assuming spherical symmetry. We present a software package that includes the calculation of a fully three-dimensional correlation function including the Coulomb interaction.

\end{abstract}

\section{Introduction}\label{s:intro}

Bose--Einstein-correlations, i.e., momentum correlations of identical bosonic particles, are an important class of observables in high-energy heavy-ion physics. Due to quantum statistical properties, momentum correlations are connected to the space-time geometry of the particle-emitting source and provide an essential tool for studying the strongly interacting Quark Gluon Plasma (QGP) created in collisions of heavy nuclei~\cite{Goldhaber:1960sf,Kopylov:1974th,Koonin:1977fh,Yano:1978gk,Lednicky:1981su,Pratt:1984su,Podgoretsky:1989bp,Pratt:1997pw,Wiedemann:1999qn}. 

In experimental studies, a usual approach is to assume a functional form for the source distribution, calculate the momentum correlation function stemming from it, and test this assumption on the measured momentum correlations~\cite{PHENIX:2004yan,STAR:2004qya,PHENIX:2024vjp,Kincses:2024sin,PHENIX:2017ino,CMS:2023xyd,NA61SHINE:2023qzr}. As detailed in our previous studies~\cite{Kincses:2019rug,Kurgyis:2020vbz,Nagy:2023zbg}, when calculating correlations of identical charged particles, it is essential to correctly take into account the final-state Coulomb interaction. This is a numerically challenging task; hence, in many previous experimental analyses, various approximations were applied. In Ref.~\cite{Nagy:2023zbg}, we developed a novel method allowing for a numerically efficient calculation of the two-particle Bose--Einstein-correlations (including the Coulomb interaction) for spherically symmetric sources. In this follow-up work, we present a generalization of this previous method to non-spherical sources. We also present a ready-to-use software package for the calculation of fully three-dimensional correlation functions~\cite{3DCoulCorrLevyIntegral}. 

The structure of this paper is as follows. In Sect.~\ref{s:BEC} we introduce the basic definitions of Bose-Einstein correlations including the Coulomb effect. In Sect.~\ref{s:integral} we discuss the new three-dimensional method of calculating the Coulomb-integral in Fourier space. Finally, in Sect.~\ref{s:results}, we discuss the results of the new calculation and its implications for experimental measurements.

\section{Bose-Einstein correlations and the Coulomb effect}\label{s:BEC}

The observable two-particle correlation function in a heavy-ion collision experiment is defined as
\begin{align}
\label{e:c2def}
\quad C_2(\bs p_1,\bs p_2)=\frac{N_2(\bs p_1,\bs p_2)}{N_1(\bs p_1)N_1(\bs p_2)},
\end{align}
where $\bs p_1$ and $\bs p_2$ are the particle momentum $\textnormal{(three-)vectors}$, $N_1(\bs p)\,{=}\,E\frac{\m dn}{\m d^3\bs p}$ is the single-particle invariant momentum distribution (the number of particles produced within a given invariant volume element in momentum space, with $E\,{=}\,\sqrt{m^2{+}p^2}$ being the particle energy corresponding to momentum $\bs p$, where $m$ is the particle mass), and $N_2(\bs p_1,\bs p_2)\,{=}\,E_1E_2\frac{\m d^2n}{\m d^3\bs p_1\,\m d^3\bs p_2}$ is the two-particle invariant momentum distribution, i.e. the number of particle \textit{pairs} produced with momenta $\bs p_1$ and $\bs p_2$.%
\footnote{
Throughout, three-vectors are denoted by boldface letters, their magnitude (modulus) by standard typeset letters. Four-vectors are also denoted by standard typeset letters; this can cause no confusion.
}
For two particles one also often uses the average momentum $\bs K$ and relative momentum $\bs Q$ variables defined as
\begin{align}
\quad \bs K = \frac{\bs p_1\,{+}\,\bs p_2}2, \qquad \bs Q = \bs p_1 \,{-}\, \bs p_2,
\end{align}
and we also introduce $\bs k = \bs Q/2$ for convenience. Also, instead of the particle production coordinates $\bs r_1$ and $\bs r_2$, one can utilize the average coordinate $\bs\rho$ and relative coordinate $\bs r$, also called pair separation, defined as
\begin{align}
\quad \bs\rho = \frac{\bs r_1\,{+}\,\bs r_2}2, \qquad \bs r = \bs r_1 \,{-}\, \bs r_2.
\end{align}
A central object of exploration in femtoscopical investigations is the spatial pair source distribution (also called relative coordinate distribution or spatial correlation function) of the produced particles (identical bosons), denoted by $D(\bs r)$ as a function of spatial separation $\bs r$. This spatial pair distribution also depends on the particle momenta (of which, owing to the so-called smoothness approximation~\cite{Pratt:1997pw}, usually only the dependence on average momentum $\bs K$ is taken into account); we suppress this dependence in our notation, since it is understood that the variois parameters of the coordinate dependence of $D(\bs r)$ may depend on the average mometum $\bs K$. Just as in Ref.~\cite{Nagy:2023zbg}, we note that in principle the particle creation also depends on time, so in turn the pair separation distribution $D$ depends also on the time separation of the particle emissions (i.e. it is not instantaneous); however, this relative time dependence can be (and usually is) factorized into the coordinate dependence by appropriate modifications of the spatial source parameters; see e.g.~Refs.~\cite{Lokos:2016fze,Cimerman:2017lmm}. So in practice we do not lose generality by treating the two-particle source function as if it depended only on the pair separation $\bs r$, and by calculating in a three-dimensional setting. We also note that if one is given the single particle source distribution $S(\bs r)$, then $D(\bs r)$ is recovered as its autoconvolution function,
\begin{align}
\quad D(\bs r) = \sint{-2pt}{-3pt}{}{}\m d\bs\rho\,S\big(\bs\rho{+}\rect2\bs r\big)S\big(\bs\rho{-}\rect2\bs r\big),
\label{e:SD}
\end{align}
but from now on, we use mainly the $D(\bs r)$ function, as it in itself has a direct connection to the observable Bose-Einstein correlations. The cornerstone of this connection is the Yano-Koonin formula~\cite{Yano:1978gk}, written up in a form suited for the utilization of $D(\bs r)$ as
\begin{align}
\quad C_2(\bs Q) = \sint{-2pt}{-3pt}{}{}\m d^3\bs r\,D(\bs r)|\psi_{\bs k}(\bs r)|^2.
\label{e:C2KP:D}
\end{align}
Here $\psi_{\bs k}(\bs r)$ is the final state pair wave-function of the produced particles. The functional form which we utilize below is the standard one that takes the final state Coulomb interaction into account; it is written up as
\begin{align}
\label{e:psikr}
\quad\psi_{\bs k}(\bs r) &= \frac{\mathcal N^*}{\sqrt2}e^{-ikr}\Big[M\big(1{-}i\eta,1,i(kr{+}\bs k\bs r)+\nonumber\\
&\qquad\qquad\qquad +M\big(1{-}i\eta,1,i(kr{-}\bs k\bs r)\big)\Big].
\end{align}
The $\eta$ quantity is the \textit{Sommerfeld parameter},
\begin{align}
\quad
\eta=\frac{mc^2\alpha}{2\hbar kc},
\end{align}
with $\alpha\,{=}\,\frac{q_e^2}{4\pi\varepsilon_0}\rec{\hbar c}\,{\approx}\,\rec{137}$ being the fine structure constant. The normalization factor $\c N$, together with its square modulus (known as the \textit{Gamow factor}) is expressed as
\begin{align}
\quad\c N = e^{-\pi\eta/2}\Gamma(1{+}i\eta)\quad\Rightarrow\quad |\c N|^2 = \frac{2\pi\eta}{e^{2\pi\eta}\,{-}\,1},
\label{e:Gamow}
\end{align}
where $\Gamma(z)$ is the Gamma function.%
\footnote{
The properties $z\Gamma(z)\,{=}\,\Gamma(z{+}1)$ and $\Gamma(z)\Gamma(1{-}z)\,{=}\,\frac\pi{\sin(\pi z)}$ lead to the expression written up for $|\c N|^2$ in \Eq{e:Gamow}.
In \Eqs{e:confhypdef}{e:hypdefpowerseries} we also use the shorthand denotation $\frac{\Gamma(a+n)}{\Gamma(a)}$ for the rising factorial,
$a(a{+}1)(a{+}2)\ldots(a{+}n{-}1)$.
}
Finally, $M(a,b,z)$ is the confluent hypergeometric function:
\begin{align}
\label{e:confhypdef}
&\quad M(a,b,z)  =  \sum_{n=0}^\infty\frac{\Gamma(a{+}n)}{\Gamma(a)}\frac{\Gamma(b)}{\Gamma(b{+}n)}\frac{z^n}{n!}.
\end{align}
In the next section, we detail our new method to calculate the integral in \Eq{e:C2KP:D}, to be able to work out the implications of various assumptions for $D(\bs r)$ and to test them against experimental data. Remarkably, methods for such effective calculations were scarce for a long time, and even many contemporary data analyses employ approximations, some of them being of various credibility. So our new method and the conclusions drawn from its application paves the way for precision analysis of Bose-Einstein correlation measurements.

Note that this next section is (together with the assorted Appendices) somewhat technical. The main result, the expression of the calculated correlation function in ~\Eq{e:C2AA} seems refreshingly simple, but the actual implementation of the calculation of its terms are rather lengthily explained; reward comes in the form of computational speed and accuracy.

\section{Coulomb integral in Fourier space}\label{s:integral}

Our method presented here relies on the $D(\bs r)$ pairwise source function to be expressed as a Fourier transform:
\begin{align}
\quad D(\bs r) = \rec{(2\pi)^3}\sint{-2pt}{-2pt}{}{}\m d^3\bs q\,f(\bs q)e^{i\bs q\bs r}.
\end{align}
Many if not all interesting cases of functional forms (e.g. Gaussian, or the Lévy form detailed in Section~\ref{s:results}) used for the modeling of $D(\bs r)$ are such that writing up or computing their Fourier transform (the corresponding $f$ function) is easy, sometimes easier than the $D(\bs r)$ function itself. For the sake of mathematical correctness of the considerations below, we place some technical assumptions on the $f(\bs q)$ function; all the physically relevant cases fulfill these. In particular, $f(\bs q)$ is assumed to be a continuous bounded integrable function.

As already explained in Ref.~\cite{Nagy:2023zbg}, we can plug this expression into the version of the Yano-Koonin formula written up with $D(\bs r)$, \Eq{e:C2KP:D}:
\begin{align}
\quad C_2(\bs Q) = \rec{(2\pi)^3}\sint{-2pt}{-3pt}{}{}\m d^3\bs r\,|\psi_{\bs k}(\bs r)|^2\sint{-2pt}{-3pt}{}{}\m d^3\bs q\,f(\bs q)e^{i\bs q\bs r},
\label{e:CDf0}
\end{align}
and we would like to proceed by interchanging the integrals over $\bs r$ and over $\bs q$ here, in order to spare the need to numerically perform first a Fourier transform (of $f$), then an ,,almost inverse Fourier transform'' (i.e. the integral over $\bs r$ with $|\psi_{\bs k}(\bs r)|^2$ as a kernel). However, as they are written up, these integrals over $\bs r$ and $\bs q$ are not interchangeable here. Instead we can first insert an $e^{-\lambda r}$ regulator into \Eq{e:CDf0}, with which we can write
\begin{align}
&\quad C_2(\bs Q)
= \sint{-2pt}{-3pt}{}{}\m d^3\bs r\lim_{\lambda\to0}e^{-\lambda r}|\psi_{\bs k}(\bs r)|^2\sint{-2pt}{-4pt}{}{}\frac{\m d^3\bs q}{(2\pi)^3}f(\bs q)e^{i\bs q\bs r}
\stackrel{1.}= \nonumber\\ &\qquad\stackrel{1.}=\lim_{\lambda\to0}\sint{-2pt}{-3pt}{}{}\m d^3\bs r\,\sint{-2pt}{-3pt}{}{}\frac{\m d^3\bs q}{(2\pi)^3}\,e^{-\lambda r}
|\psi_{\bs k}(\bs r)|^2\,f(\bs q)e^{i\bs q\bs r} \stackrel{2.}= \nonumber\\
&\qquad\stackrel{2.}=\lim_{\lambda\to0}\sint{-2pt}{-3pt}{}{}\frac{\m d^3\bs q}{(2\pi)^3}\,f(\bs q)
\sint{-2pt}{-3pt}{}{}\m d^3\bs r\,e^{-\lambda r}|\psi_{\bs k}(\bs r)|^2\,e^{i\bs q\bs r}.
\label{e:CDf1}
\end{align}
These steps are a recapitulation of Eq.~(39) in Ref.~\cite{Nagy:2023zbg}; in that paper (in Appendix A) we also highlighted the mathematical theorems and notions about (Lebesgue) integrability that our new calculational method makes heavy use of. We repeat these here as well:%
\footnote{
Again, for a detailed discussion, see any of the standard textbooks on mathematical analysis, such as Ref.~\cite{Rudin}.
}
\vspace{2mm}\\
$\bullet$
A real or complex valued function $F$ is integrable if and only if $|F|$ is integrable. If there is a $G$ function for which $|G(x)|\,{\ge}\,|F(x)|$ for almost all $x$ (with $x$ being a one-dimensional or multi-dimensional real integration variable), and $G$ is integrable, then so is $F$.
\vspace{2mm}\\
$\bullet$
\textit{Lebesgue's theorem} makes interchanging integrals and limits justified in a wide variety of cases. Let $F_\lambda(x)$ be integrable functions, with $\lambda$ as a parameter, and for almost all $x$ let $F(x)\,{:=}\,\lim_\lambda F_\lambda(x)$ exist (for any reasonable type of limit). If there is a ($\lambda$-independent) $G$ function for which for almost all $x$, $|G(x)|\,{\ge}\,|F_\lambda(x)|$, for any $\lambda$, then the integrals of the $F_\lambda$ functions (the $\int F_\lambda$ values) converge, the pointwise limiting function $F$ is integrable, and $\lim_\lambda \int F_\lambda = \int F$, i.e. the limit and the integral are interchangeable. (In cases when there is no such ,,dominant'' $G$ function, none of these statements are necessarily true.)
\vspace{2mm}\\
$\bullet$
Fubini's theorem concerns (the interchange of) multiple integrals. The main point is that if an $F(x,y)$ function is such that its modulus is integrable in one order; i.e. if the $\int\m dx\big(\int\m dy\,|F(x,y)|\big)$ integral exists, then $F$ itself is integrable in both orders, and these integrals coincide: $\int\m dx\int\m dy\,F(x,y)\,{=}\,\int\m dy\int\m dx\,F(x,y)$, i.e. the integrals are interchangeable.
\vspace{2mm}\\
\indent
Based on these, one may verify that the steps in \Eq{e:CDf1} are justified: the exchange of the limit and the integral in Step 1 by virtue of the Lebesgue theorem (with $|\psi_{\bs k}(\bs r)|^2D(\bs r)$ being the dominant function, even if $D(\bs r)$ is written up as a Fourier transform), then Step 2 by virtue of the Fubini theorem; the modulus being $|f(\bs q)|\cdot e^{-\lambda r}\cdot|\psi_{\bs k}(\bs r)|^2$). However, in the last form the limit and the integral is not interchangeable.

We may now substitute the pair wave-function from \Eq{e:psikr} into \Eq{e:CDf1}, yielding four terms. In two of these, one can apply an $\bs r\to-\bs r$ substitution, with which these terms will be identical to two others, respectively. We will be thus left with two terms as
\begin{align}
&\quad\sint{-2pt}{-3pt}{}{}\m d^3\bs r\,e^{-\lambda r+i\bs q\bs r}|\psi_{\bs k}(\bs r)|^2 = \nonumber\\
&\qquad\qquad\qquad = |\c N|^2\cdot\Big[\c D_{1\lambda}(\bs q)\,{+}\,\c D_{2\lambda}(\bs q)\Big],
\end{align}
where $\c D_{1\lambda}(\bs q)$ and $\c D_{2\lambda}(\bs q)$ are defined as
\begin{align}
& \quad \c D_{1\lambda}(\bs q) =\slint{-3pt}{2pt}{}\m d^3\bs r\,e^{-\lambda r}M\big(1{-}i\eta,1,i(kr{+}\bs k\bs r)\big)\times\nonumber\\
& \qquad\qquad\qquad\quad \times M\big(1{+}i\eta,1,-i(kr{+}\bs k\bs r)\big)e^{i\bs q\bs r},
\label{e:D1lambdadef}\\
& \quad \c D_{2\lambda}(\bs q) =\slint{-3pt}{2pt}{}\m d^3\bs r\,e^{-\lambda r}M\big(1{-}i\eta,1,i(kr{+}\bs k\bs r)\big)\times\nonumber\\
& \qquad\qquad\qquad\quad \times M\big(1{+}i\eta,1,-i(kr{-}\bs k\bs r)\big)e^{i\bs q\bs r}.
\label{e:D2lambdadef}
\end{align}
So our task is to calculate the correlation function as
\begin{align}
\quad C(\bs Q)\,{=}\,\frac{|\c N|^2}{8\pi^3}\hspace{-2pt}\lim_{\lambda\to0}\slint{-1pt}{1pt}{}\m d^3\bs q\,f(\bs q)\Big[\c D_{1\lambda}(\bs q)\,{+}\,\c D_{2\lambda}(\bs q)\Big].
\label{e:CQD}
\end{align}
First the calculation of $\c D_{1\lambda}(\bs q)$ and $\c D_{2\lambda}(\bs q)$ for finite $\lambda{>}0$ are performed, just as in \cite{Nagy:2023zbg}, by utilizing a
slightly updated and modified method by Nordsieck~\cite{Nordsieck:1954zz}; for details, see~\ref{s:app:D12}. The result can be expressed using a wealth of auxiliary
quantities; most importantly the kinematic variables
\begin{align}
\quad
\bs q_\pm = \bs q \pm 2\bs k.
\end{align}
Remember, $\bs Q\,{=}\,2\bs k$ throughout. We will make extensive use of the hypergeometric function, defined for complex $z\,{\in}\,\B C$ variable and $a,b,c\,{\in}\,\B C$
parameters as 
\begin{align}
&\quad {}_2F_1\big(a,b,c,z\big)\,{=}\sum_{n=0}^\infty\frac{\Gamma(a{+}n)\Gamma(b{+}n)}{\Gamma(a)\Gamma(b)}\frac{\Gamma(c)}{\Gamma(c{+}n)}\frac{z^n}{n!}
\label{e:hypdefpowerseries}
\end{align}
when $|z|{<}1$, and by analytic continuation elsewhere. The following parameter combinations will be used:
\begin{align}
\label{e:Fplusdef}
\quad&\c F_+(x) := {}_2F_1(i\eta,1{+}i\eta,1,x),\\
&\c F_-(x):=(1{+}i\eta)\cdot{}_2F_1({-}i\eta,1{-}i\eta,2,x).
\label{e:Fminusdef}
\end{align}
We use the variables
\begin{align}
\quad
x_1 = \frac{4(\bs q\bs k{-}ik\lambda)^2}{(\lambda^2{+}q^2)^2},\qquad
x_2 = \frac{4[k^2q^2{-}(\bs q\bs k)^2]}{(\lambda^2{+}q^2)(\lambda^2{+}q_+^2)},
\label{e:x1x2def}
\end{align}
and the auxiliary quantities
\begin{align}
\label{e:U1def}
\quad &\c U_{1\pm}=\frac{(\lambda^2{+}q^2)^{\pm 2i\eta}}{\z{\lambda^2{+}\bs q\bs q_\mp{\pm}2ik\lambda}^{\pm 2i\eta}},\\
&\c U_{2\pm}=\frac{((\lambda^2{+}q^2)(\lambda^2{+}q_+^2))^{\pm i\eta}}{\z{\lambda^2{+}\bs q\bs q_+{\pm}2ik\lambda}^{\pm 2i\eta}},
\label{e:U2def}
\end{align}
with which the expression of $\c D_{1\lambda}(\bs q)$ and $\c D_{2\lambda}(\bs q)$ are then most easily written up as derivatives:
\begin{align}
\label{e:D1I1main}
\quad& \c D_{1\lambda}(\bs q) = -4\pi\td{}\lambda\frac{\c U_{1-}\c F_+(x_1)}{\lambda^2{+}q^2},\\
& \c D_{2\lambda}(\bs q) = -4\pi\td{}\lambda\frac{\c U_{2-}\c F_+(x_2)}{\lambda^2{+}q_+^2}.
\label{e:D2I2main}
\end{align}
More explicit expressions of $\c D_{1\lambda}(\bs q)$ and $\c D_{2\lambda}(\bs q)$ are given in \Eqs{e:app:D1lambda}{e:app:D2lambda} in \ref{s:app:D12}.

The next step is to bring the result of the $\lambda\,{\to}\,0$ limit in our expression of the correlation function, \Eq{e:CQD} into an explicit form with the limit already ,,performed''; meaning that in the final expression there is no need to take limits numerically. As a preliminary note, one can observe that if $\eta{=}0$, i.e. with the Coulomb interaction turned off/neglected, $\c U_{1\pm}\,{=}\,\c U_{2\pm}\,{=}\,\c F_+(x)\,{=}\,1$, so in this case (denoted by an ,,$\eta{=}0$'' label in the superscript) we have simply
\begin{align}
&\quad \c D^{(\eta=0)}_{1\lambda} = \frac{8\pi\lambda}{(\lambda^2{+}q^2)^2},\qquad 
\c D^{(\eta=0)}_{2\lambda} = \frac{8\pi\lambda}{(\lambda^2{+}q_+^2)^2},
\label{e:Deta0}
\end{align}
and the $\lambda\,{\to}\,0$ limit of these expressions is a standard approximation of three-dimensional Dirac deltas:
\begin{align}
&\quad\lim_{\lambda\to0}\c D^{(0)}_{1\lambda}(\bs q) = 8\pi^3\,\delta^{(3)}(\bs q),\\
&\quad\lim_{\lambda\to0}\c D^{(0)}_{2\lambda}(\bs q) = 8\pi^3\,\delta^{(3)}(\bs q{+}2\bs k).
\end{align}
The meaning of these relations is precisely that together with these, \Eq{e:CQD} implies  
\begin{align}
\quad C^{(\eta=0)}(\bs Q) = f(\bs 0) \,{+}\, f({-}2\bs k) = 1 + f(\bs Q), 
\label{e:Cfreecheck}
\end{align}
recalling that $\bs Q\,{=}\,2\bs k$, and also that we assumed $f$ to be an even function with $f(\bs 0)\,{=}\,1$, and finally that if $\eta{=}0$, $|\c N|^2\,{=}\,1$. This is
indeed the standard result for the interaction-free case: the correlation is given by the Fouirier transform of the pair source function.%
\footnote{
If we express the two-particle source function $D(\bs r)$ with the single-particle source function $S(\v r)$ as in \Eq{e:SD} then this implies the somewhat more familiarly sounding conclusion that the correlation function is the modulus square of the Fourier transform of the one-particle source function $S(\v r)$.
} 

We now return to the $\eta\,{\neq}\,0$ case. After a detailed investigation of the $\lambda\,{\to}\,0$ limit in \Eq{e:CQD}, one arrives at an expression of $C(\bs Q)$ in the desired form. We leave the majority of this derivation to \ref{s:app:D12limit}. The result can be written up as
\begin{align}
\quad C(\bs Q) = |\c N|^2\Big[1 \,{+}\, f(\bs Q) \,{+}\, \frac\eta\pi\big[\c A_1\,{+}\,\c A_2\big]\Big],
\label{e:C2AA}
\end{align}
a direct generalization of the similar result in the spherically symmetric case, see Eq.~(48) in Ref.~\cite{Nagy:2023zbg}. The first term in the parenthesis (unity) is $f(\bs 0)$, which is equal to 1 in our normalization. The terms denoted by $\c A_1$ and $\c A_2$ (stemming from the limiting cases of $\c D_{1\lambda}$ and $\c D_{2\lambda}$, respectively) depend on ${\bs Q}$ and the $f$ function itself in a complicated manner, as detailed below.

For writing up the explicit expressions of $\c A_1$ and $\c A_2$, it turns out to be worthwhile to introduce two new convenient coordinate systems in the $\bs q$-space (the space of the integration variable), one for $\c A_1$ and another for $\c A_2$. We introduce these new coordinates below, but already note here that they are illustrated in Figs.~\ref{f:A1coord}-\ref{f:A2coord}. These are also fixed with respect to the $\bs Q$ vector, the final variable of the $C(\bs Q)$ correlation function. In the following, we denote by $q_\parallel$ the component of the $\bs q$ vector parallel to $\bs Q$, by $q_\perp$ the component perpendicular to it, and we introduce a $\varphi$ azimuthal as a new coordinate, that is measured around the direction of $\bs Q$. The unit vector in the $\bs Q$ direction is denoted by $\bs e_\parallel$; its polar and azimuthal angles by $\Theta$ and $\Phi$, so
\begin{align}
\quad \bs Q \,{=}\,Q\bs e_\parallel,\qquad\tn{where}\quad \bs e_\parallel = \begin{pmatrix} \sin\Theta\cos\Phi \\ \sin\Theta\sin\Phi \\ \cos\Theta \end{pmatrix}.
\end{align}
In the plane perpendicular to $\bs Q$, we are free to choose two $\bs e_{\perp1}$ and $\bs e_{\perp2}$ basis vectors; for instance,
\begin{align}
\quad \bs e_{\perp1} = \begin{pmatrix} \cos\Theta\cos\Phi \\ \cos\Theta\sin\Phi \\ {-}\m{sin}\,\Theta \end{pmatrix},\quad
\bs e_{\perp2} = \begin{pmatrix} {-}\m{sin}\,\Phi \\ \cos\Phi \\ 0 \end{pmatrix},
\end{align}
with which the $\bs q$ vector is written up as
\begin{align}
\label{e:qnewcoord}
\quad \bs q = q_\parallel\cdot\bs e_\parallel \,{+}\, q_{\perp1}{\cdot}\,\bs e_{\perp1} \,{+}\,q_{\perp2}{\cdot}\,\bs e_{\perp2},
\end{align}
where $q_{\perp1} = q_\perp{\cdot}\,\cos\varphi$ and $q_{\perp2} = q_\perp{\cdot}\,\sin\varphi$. To define the new coordinate systems for the calculation of $\c A_1$ and
$\c A_2$, in each case we thus have to specify two new coordinates that express $q_\parallel$ and $q_\perp$; the third coordinate will then be the $\varphi$ azimuthal angle
in both cases.

In case of $\c A_1$, the convenient new coordinates are denoted by $a$ and $\beta$, with domains
\begin{align}
\quad a\in\B R,\quad \beta\in\B R_0^+\quad\tn{(and $\varphi\in\;\;]{-}\pi,\pi])$}.
\label{e:A1coord:dom}
\end{align}
In the $q_\parallel$--$q_{\perp1}$ plane, the level lines of $a$ are circles that all go through the $q_\parallel\,{=}\,q_{\perp1}{=}\,0$ origin, and all have their centers on the axis of the $\bs e_\parallel$ vector; the values of $a$ are chosen in a way that of such circles, the one with a given $a$ intersects the $\bs e_\parallel$ axis at a (signed) distance of $ka$ from the origin. For example, the $a\,{=}\,{\pm}1$ circles go through the endpoints of the $\pm\bs Q\,{=}\,{\pm}2\bs k$ vectors. The level lines of the $\beta$ coordinate are orthogonal to these circles; they are also circles with all of them passing through the origin and having their centers on the $\bs e_{\perp1}$ axis; they are obtained from the $a\,{\ge}\,0$ circles with a rotation by 90$^\circ$. The $a$ and $\beta$ coordinates thus cover the $q_{\perp1}{\ge}\,0$ half plane; the $\varphi$ azimuthal angle serves to sweep the whole $\bs q$ space. The expression of the $\bs q$ components in this new coordinate system is
\begin{align}
\quad \begin{pmatrix} q_{\perp1} \\ q_{\perp2} \\ q_\parallel \end{pmatrix}
= \fracd{2k\,a\beta}{a^2{+}\beta^2}\begin{pmatrix} a\cos\varphi \\ a\sin\varphi \\ \beta \end{pmatrix}.
\label{e:A1coord}
\end{align}
In case of $\c A_2$, the two new coordinates are denoted by $b$ and $y$, with domains
\begin{align}
\quad b\in\B R_0^+,\quad y\in[{-}1,1]\quad\tn{(and $\varphi\in\;\;]{-}\pi,\pi])$}.
\label{e:A2coord:dom}
\end{align}
The level lines of $y$ in the $q_\parallel$--$q_{\perp1}$ plane are the Apollonian circles of the points $\bs q\,{=}\,{-}\bs Q$ and $\bs q{=}\bs 0$ (the loci of points with fixed ratio of distances to these two points); a given $y\,{\in}\,[-1,1]$ value specifies such a cirlce by having its intersection with the $[{-}\bs Q,\bs 0]$ segment at point ${-}\frac{1{-}y}2\bs Q$; in this way the circles with $y\,{>}\,0$ encircle the $\bs q\,{=}\,0$ point, those with $y\,{<}\,0$ encircle the $\bs q\,{=}\,{-}\bs Q$ point, while the $y\,{=}\,0$ level line is an infinte straight line, the perpendicular bisector of the $[{-}\bs Q,\bs 0]$ segment. The level lines of $b$ are circles orthogonal to the $y$ level lines; these are circles that for any $b$ go through both the $\bs q\,{=}\,\bs 0$ origin and the $\bs q\,{=}\,{-}\bs Q$ point; out of these the one that corresponds to a given $b$ value has its intersection with the $y\,{=}\,0$ straight line at a distance of $k\sqrt b$ in the positive $q_{\perp1}$ direction. In this way the $b$--$y$ coordinates cover the $q_{\perp1}{\ge}\,0$ half plane; again the inclusion of the $\varphi$ azimuthal angle serves to cover the whole $\bs q$ space. The components of $\bs q$ in this coordinate system are
\begin{align}
\quad \begin{pmatrix} q_{\perp1} \\ q_{\perp2} \\ q_\parallel \end{pmatrix}
= k\frac{1{-}y}{1{+}by^2}\begin{pmatrix} \sqrt b(1{+}y)\cos\varphi \\ \sqrt b(1{+}y)\sin\varphi \\ by-1 \end{pmatrix}.
\label{e:A2coord}
\end{align}
Figs.~\ref{f:A1coord} and \ref{f:A2coord} illustrate these new coordinate systems.

We introduce the notations $f_1$ and $f_2$ as the co\-ordi\-nate-aligned versions of the $f(\bs q)$ function, defined as
\begin{align}
&\quad f_1(a,\beta,\varphi) = f\big(\bs q(a,\beta,\varphi)\big) - f(\bs q{=}\bs 0),\\
&\quad f_2(b,y,\varphi) = f\big(\bs q(b,y,\varphi)\big) - f(\bs q\,{=}\,{-}\bs Q),
\end{align}
with $\bs q(a,\beta,\varphi)$ and $\bs q(b,y,\varphi)$ being understood as the $\bs q$ values expressed as functions of the new coordinates $a$, $\beta$, $\varphi$ as in
\Eq{e:A1coord}, and as functions of $b$, $y$, $\varphi$ as in \Eq{e:A2coord}, respectively. With these, the expression of the $\c A_1$ term in \Eq{e:C2AA} is
\begin{align}
\label{e:A1final}
&\quad\pi\c A_1\,{=}\,
\sint{-2pt}{-13pt}{{-}\infty}\infty\m da\sint{-0pt}{-12pt}0\infty\m d\beta\sint{-2pt}{-10pt}{-\pi}\pi\m d\varphi\,\frac{aY({-}a)}{\beta(a{+}1)}\,\m{Re}\Bigg\{
\frac{|a|^{2i\eta}}{|a{+}1|^{2i\eta}}\times\nonumber\\
&\;\,{\times}\bigg[G^*\frac{2i\,f_1({-}1,\beta,\varphi)}{(\beta^2{+}1)(a{-}i)^2}\,{-}\,
\c F_-\big(\fract{1{+}a{\cdot}i0}{a^2}\big)\frac{f_1(a,\beta,\varphi)}{a^2{+}\beta^2}\bigg]\hspace{-1pt}\Bigg\}.
\end{align}
The $\c A_2$ term is a sum of two terms of different nature,  
\begin{align}
\qquad \c A_2 = \c A_{2\c P} + \c A_{2\delta},
\end{align}
where the $\c A_{2\c P}$ term is a three-dimensional integral, 
\begin{align}
&\quad\pi\c A_{2\c P}\,{=}\,
\sint{-2pt}{-12pt}0\infty\m db\sint{-2pt}{-10pt}{-1}1\m dy\sint{-0pt}{-10pt}{-\pi}\pi\m d\varphi\,
\frac{1{-}y}{1{+}y}\frac{Y(b)}{b{-}1}\,\m{Re}\Bigg\{\frac{(b{+}1)^{2i\eta}}{|b{-}1|^{2i\eta}}\times\nonumber\\&\;\,{\times}\bigg[
G^*\frac{2f_2(1,y,\varphi)}{(1{+}b)(1{+}y^2)} - \c F_-\big(\fract{4b}{(1{+}b)^2}\big)\frac{f_2(b,y,\varphi)}{1{+}by^2}\bigg]\hspace{-1pt}\Bigg\},
\label{e:A2Pfinal}
\end{align}
while the $\c A_{2\delta}$ term is two-dimensional; the support of the integrand here is the surface of the $b{=}1$ sphere:
\begin{align}
&\c A_{2\delta} = \m{Im}(G)\frac{e^{2\pi\eta}{-}1}{4\pi\eta}\slint{-2pt}{-9pt}{S^2}\m d\Omega(\bs n)\,\frac{f(\bs k{-}k\bs n){-}f(2\bs k)}{1{+}\hat{\bs k}\bs n}.
\label{e:A2dfinal}
\end{align}
In these expressions, $G$ is a parameter and $Y$ is a step-like function, both connected to the analytic properties of $\c F_-(z)$, and its behavior at its
$z{=}1$ singular point: 
\begin{align}
&\quad G = 2\frac{\Gamma({-}2i\eta)}{\Gamma({-}i\eta)\Gamma(1{-}i\eta)},\label{e:Gdef}\\
&\quad Y(\xi) = \vphantom{\Bigg(}\bigg\{ 
\begin{array}{l}
e^{2\pi\eta},\;\tn{if $0\,{\le}\,\xi\,{\le}\,1$,}\\
1\quad\tn{otherwise.}
\end{array}
\end{align}
An implementation of these calculations is provided in Ref.~\cite{3DCoulCorrLevyIntegral}.

\begin{figure}
\centerline{
\includegraphics[width=0.75\linewidth]{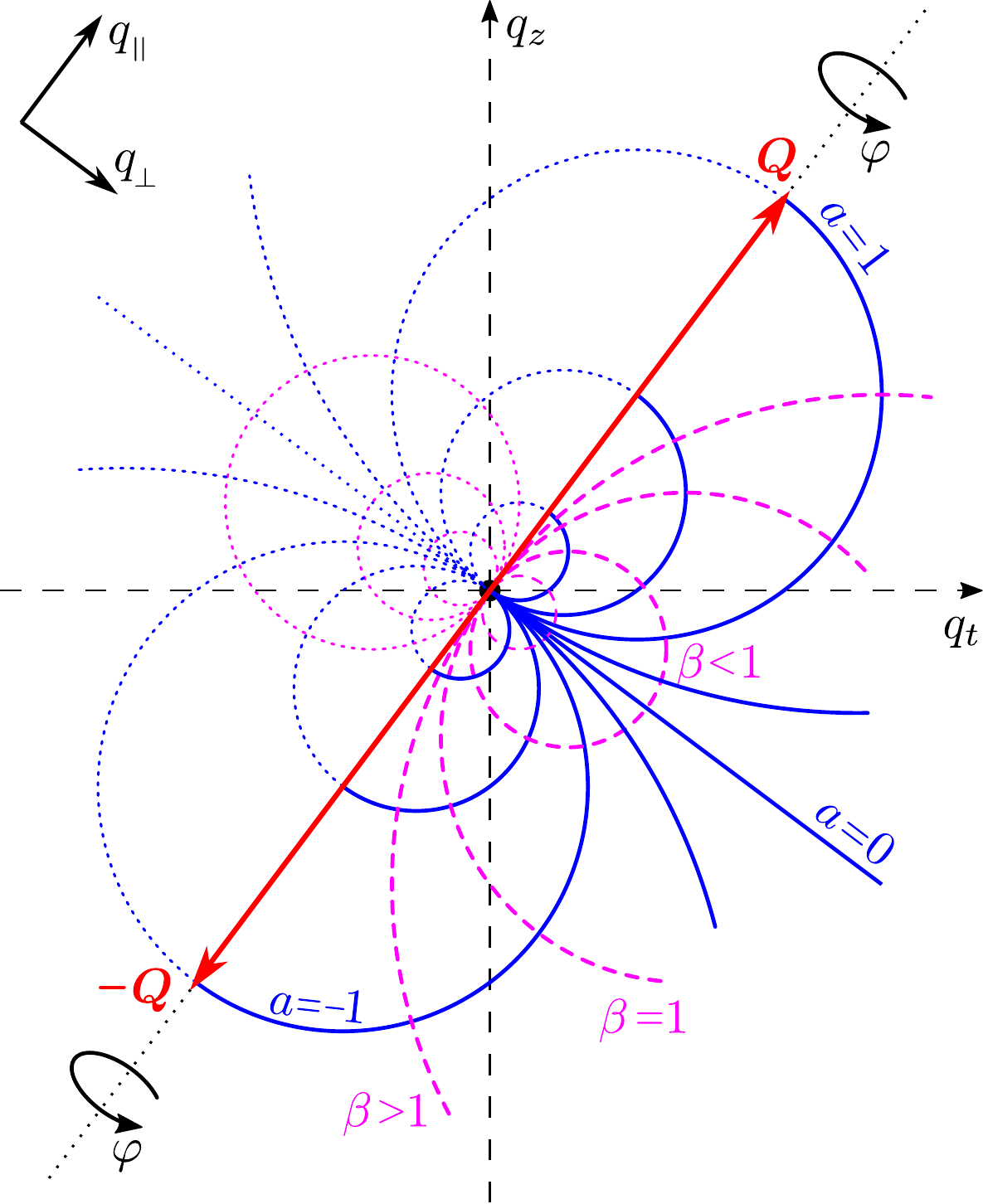}}
\caption{The $a$--$\beta$--$\varphi$ coordinate system, suited for the calculation of $\c A_1$ as in \Eq{e:A1final}.
}
\label{f:A1coord}
\end{figure}
\begin{figure}
\centerline{
\includegraphics[width=0.75\linewidth]{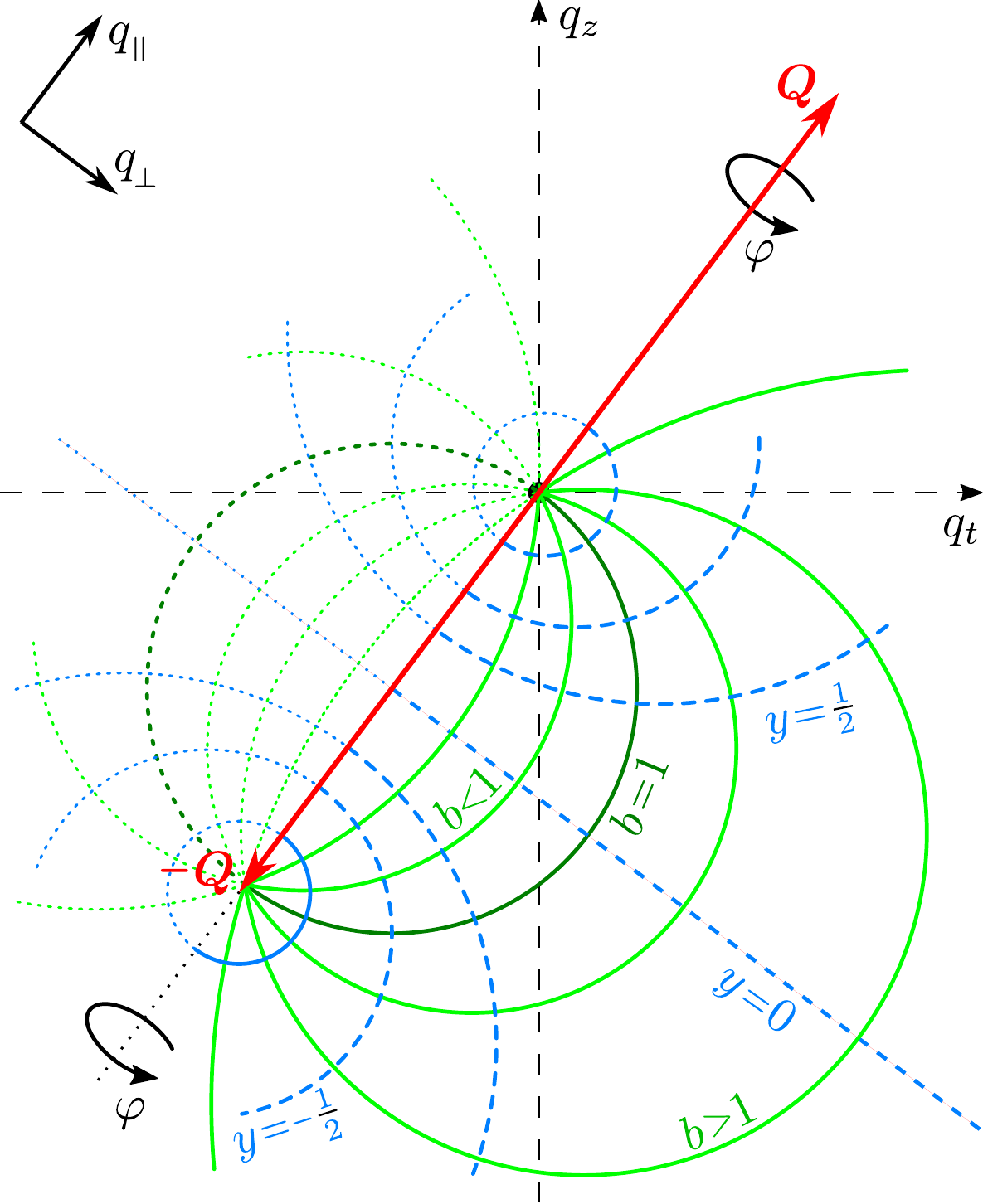}}
\caption{The $b$--$y$--$\varphi$ coordinate system, suited for the calculation of $\c A_2$ as in \Eq{e:A2Pfinal}.
}
\label{f:A2coord}
\end{figure}

\section{Results and discussion}\label{s:results}

Above we have derived the formulas for calculating Bose-Einstein correlation functions including the Coulomb final state interaction with a 3D non-spherical source function that can be expressed as a Fourier transform. Recent experimental measurements~\cite{PHENIX:2017ino,PHENIX:2024vjp,NA61SHINE:2023qzr,CMS:2023xyd,Kincses:2024sin} and phenomenological investigations~\cite{Korodi:2022ohn,Csanad:2024hva,Kincses:2024lnv,Csanad:2024jpy,Huang:2025edi,Kincses:2025izu,Kincses:2025iaf} have shown that elliptically contoured Lévy-stable distributions can provide an adequate description of the two-pion source in heavy-ion collisions. Thus, from this point on, we implement such a source shape in our calculations. In this case, the $f(\bs q)$ kernel is defined as 
\begin{align}
\quad f(\bs q) = e^{-|\bs q\bs R^2\bs q|^{\alpha/2}},
\end{align}
where $\alpha$ is called the L\'evy exponent parameter, and $\bs R^2$ is a 3 by 3 matrix, containing the six independent Lévy scale parameters. This scale parameter matrix is often approximated as diagonal (for azimuthally averaged analyses), and taken in the $out-side-long$ coordinate system (also called Bertsch-Pratt coordinate system~\cite{Bertsch:1988db,Pratt:1986ev}), where $out$ is the direction of the average transverse momentum of the particle pair, $long$ is the beam direction, and the $side$ direction is perpendicular to the two other ones.

\begin{figure*}
\includegraphics[width=0.99\linewidth]{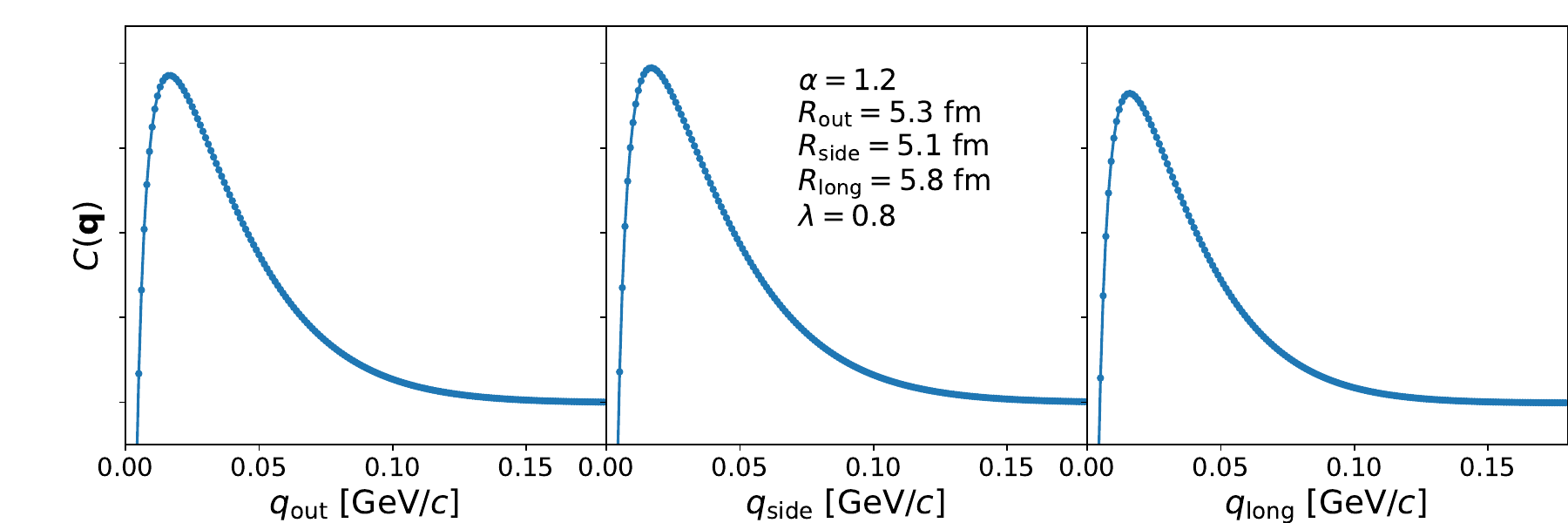}
\caption{An example plot showing the 3D correlation function of Eq.~\eqref{e:C2Coulomblambda}, with a given set of parameters, in 1D slices along the three axes of the Bertsch-Pratt coordinate system.}
\label{f:coulcorrtest}
\end{figure*}

\begin{figure*}
\includegraphics[width=0.99\linewidth]{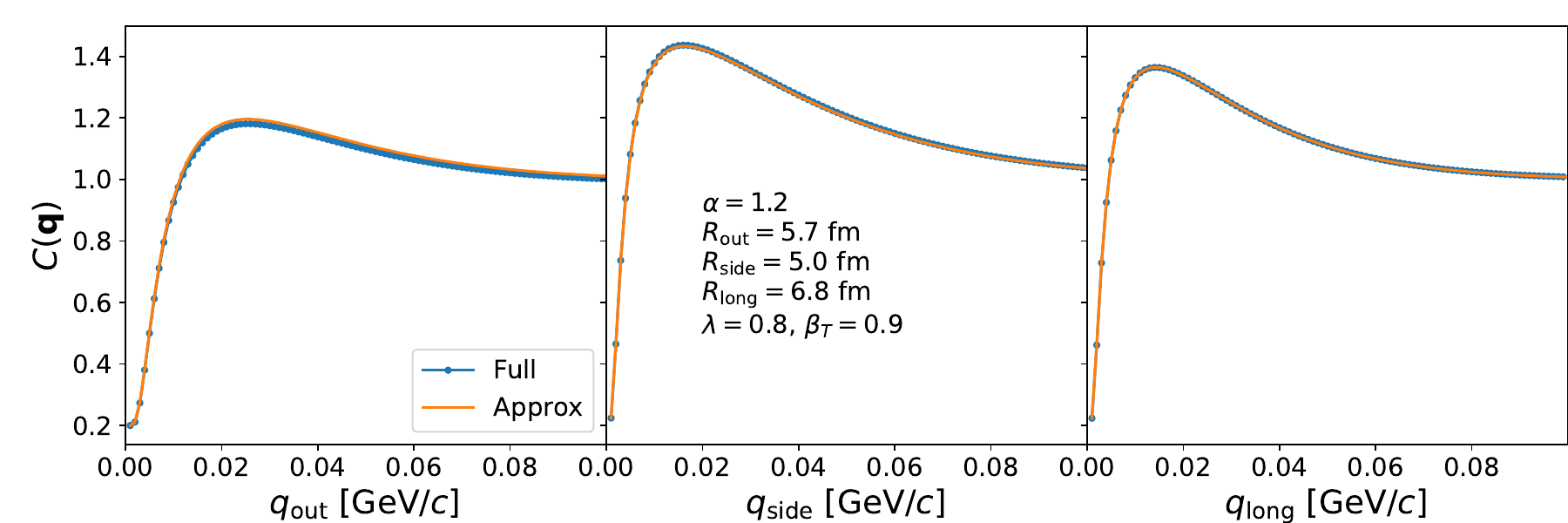}
\caption{Comparison of the full (blue markers and line) and the approximative (orange line) 3D calculation of Eqs.~\eqref{e:C2Coulomblambda} and~\eqref{e:C2Coulombapprox}, taking the LCMS-PCMS difference into account, with a given set of parameters and $\beta_T$, in 1D slices along the three axes of the Bertsch-Pratt coordinate system.}
\label{f:coulcorrcompare}
\end{figure*}

To illustrate our calculations, in Fig.~\ref{f:coulcorrtest} we show such a correlation function with a set of example parameters, in 1D slices along three axes, utilizing the $q_\mathrm{out}$--$q_\mathrm{side}$--$q_\mathrm{long}$ coordinate system. Note that in case of experimental data, similar slices (or rather: projections) are usually taken in a finite width slab or cylinder around the given axes, which smears the behavior of $C(q_i)$ ($i=$out,side,long) at around $q_i\to0$.

Correlation functions based on non-spherical sources were mostly calculated in the past by approximating the Coulomb part of the correlations with a spherically symmetric source~\cite{STAR:2003ytv,STAR:2004qya,PHENIX:2004yan,STAR:2009fks,STAR:2014shf,PHENIX:2014pnh}. This approximation can be understood if we separate the ``quantum-statistical'' and the ``final-state interaction'' part of the correlations starting from Eq.~\eqref{e:C2KP:D} as
\begin{align}
\quad C_2(\bs Q) & = \sint{-2pt}{-3pt}{}{}\m d^3\bs r\,D(\bs r)|\psi_{\bs k}(\bs r)|^2=\nonumber\\
&=K_C(\bs Q)\cdot C_2^{(0)}(\bs Q),
\end{align}
or taking into account the $\lambda$ intercept parameter~\cite{Alt:1999cs,Csorgo:1994in,PHENIX:2017ino} (that in the core-halo picture of the source function gives the relative contribution of the core as $\sqrt\lambda$),
\begin{align}
\quad C_2(\bs Q) = 1\,{-}\,\lambda \,{+}\, \lambda \cdot K_C(\bs Q)\cdot C_2^{(0)}(\bs Q).
\label{e:C2Coulomblambda}
\end{align}
Here $K_C(\bs Q)$ is the so-called Coulomb correction,
\begin{align}
\quad K_C(\bs Q) &= \frac{\sint{-2pt}{-3pt}{}{}\m d^3\bs r\,D(\bs r)|\psi_{\bs k}(\bs r)|^2}{\sint{-2pt}{-3pt}{}{}\m d^3\bs r\,D(\bs r)|\psi_{\bs k}^{(0)}(\bs r)|^2},
\label{e:C2Coulomb}
\end{align}
while $C_2^{(0)}(\bs Q)$ is the correlation function that would arise if the final state interactions of the produced particles (and $\lambda$ being less than 1) were neglected,
\begin{align}
\quad C_2^{(0)}(\bs Q) &= \sint{-2pt}{-3pt}{}{}\m d^3\bs r\,D(\bs r)|\psi_{\bs k}^{(0)}(\bs r)|^2,
\end{align}
with the modulus square of the free-particle $\psi_{\bs k}^{(0)}(\bs r)$ pair wave-function being $|\psi_{\bs k}^{(0)}(\bs r)|^2\,{=}\,1\,{+}\,\cos(\bs Q\bs r)$. The cosine term here leads precisely to the result (recalled in \Eq{e:Cfreecheck} above) that $C^{(0)}(\bs Q)$ is essentially the Fourier transform of the pair source function. So $C_2^{(0)}(\bs Q)$ is straightforward to calculate, even for a not spherically symmetric source as well.

Let us assume that $D(\bs r)$ takes the form of a 3D Lévy distribution with exponent $\alpha$ and scale matrix 
\begin{align}
\quad \bs R^2 = {\rm diag}\left(R^2_{\rm out}, R^2_{\rm side}, R^2_{\rm long}\right);    
\end{align}
then one obtains
\begin{align}
\quad C_2^{(0)}(\bs Q) = 1 + \exp\left[-(\bs Q^T \bs R^2 \bs Q)^{\alpha/2}\right].
\end{align}
However, for $C_2(\bs Q)$ one needs the calculations discussed in this paper, which are much more ``expensive'' in terms of computational time. This is not always feasible when fitting experimental data, so one often resorts to~\cite{STAR:2003ytv,STAR:2004qya,PHENIX:2004yan,STAR:2009fks,STAR:2014shf,PHENIX:2014pnh} calculating $K_C(\bs Q)$ for a spherically symmetric source, and then fitting the source parameters ($\alpha, \bs R^2, \lambda$) only in the remaining part of Eq.~\eqref{e:C2Coulomblambda}. We test the accuracy of this calculation by comparing the two cases: when one performs the full 3D calculation as in Eq.~\eqref{e:C2Coulomblambda}, and the approximation where the $K_C(\bs Q)$ part is calculated for a spherically symmetric source, taking its scale parameters as an average of the directional radii $R^2_{\rm out}, R^2_{\rm side}, R^2_{\rm long}$. For this average, one has to note that the Coulomb calculations outlined in this paper are only valid in the pair rest frame, often called Pair Co-Moving System, or PCMS. However, the source shows an approximate spherical symmetry in the Longitudunally Co-Moving Frame (LCMS)~\cite{PHENIX:2017ino}. Thus in practice to calculate $C_2(\bs Q)$ in the LCMS, one has to boost the source into PCMS, and then perform the calculation there. For this one has to know the relative velocity of the two frames, the transverse velocity $\beta_T$ (a dimensionless quantity, scaled by the speed of light). The average spherical radius in the PCMS can then be calculated as~\cite{Kurgyis:2020vbz}
\begin{align}
    \quad R_{\rm PCMS} = \sqrt{\frac{R_{\rm out}^2/(1{-}\beta_T^2) \,{+}\, R_{\rm side}^2 \,{+}\, R_{\rm long}^2}{3}}.
\end{align}
The approximation is then:
\begin{align}
\quad C_2(\bs Q) &= 1-\lambda + \nonumber\\
&+\lambda \cdot K_C^{\rm(sph. symm.)}(R_{\rm PCMS},\bs Q)\cdot C_2^{(0)}(\bs Q).
\label{e:C2Coulombapprox}
\end{align}
It turns out that if the source is spherically symmetric, then the Coulomb correction is also spherically symmetric, and it depends only on the momentum difference in the PCMS frame,
\begin{align}
\quad Q_{\rm PCMS}=\sqrt{Q_{\rm out}^2(1{-}\beta_T^2) \,{+}\, Q_{\rm side}^2 \,{+}\, Q_{\rm long}^2}.
\end{align}
This comparison is shown in Fig.~\ref{f:coulcorrcompare}, for an example set of parameters, with $\beta_T = 0.9$. It is clear that the difference is zero if the source is spherical in the PCMS, or if it spherical in the LCMS and $\beta_T=0$. For a source spherical in the LCMS, the difference grows as $\beta_T$ increases. This is illustrated in Fig.~\ref{f:coulcorrcompare.betatest}. While the difference shown there is taken as the average difference in all bins, it is important to see where the two calculations differ the most. This is shown in Fig.~\ref{f:coulcorrcompare.2D.Cdiff}. One may conclude from this that the approximation works well in many cases, although for precise data (arising from large statistics datasets) it is advised to investigate the more accurate calculation as part of systematic uncertainties of the experimental results.

\begin{figure}
\includegraphics[width=0.99\linewidth]{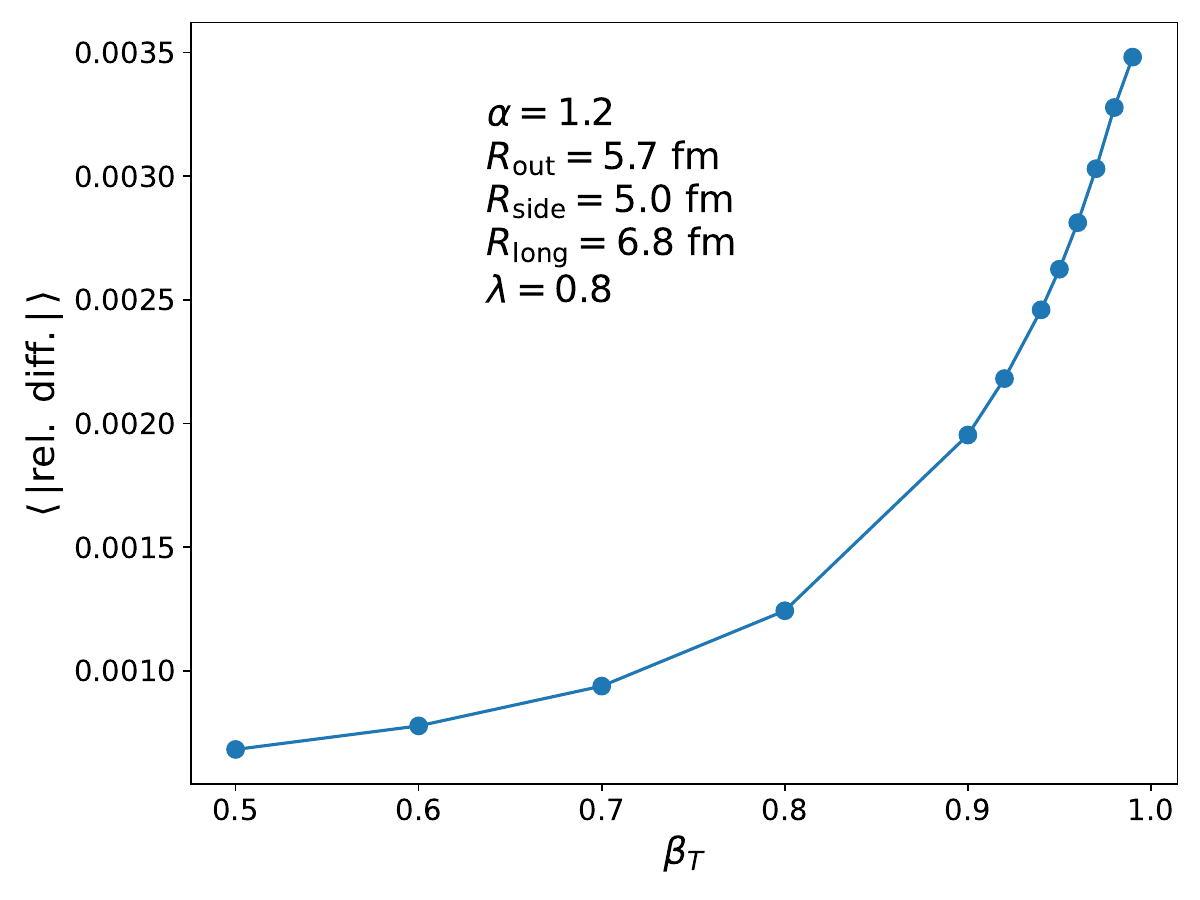}
\caption{Averaged relative difference of the full and approximative 3D calculations of Fig.~\ref{f:coulcorrcompare}, as a function of $\beta_T$. The average was taken over momentum differences with all coordinates smaller than 150 MeV/$c$.}
\label{f:coulcorrcompare.betatest}
\end{figure}



\begin{figure*}
\centering\includegraphics[width=0.98\linewidth]{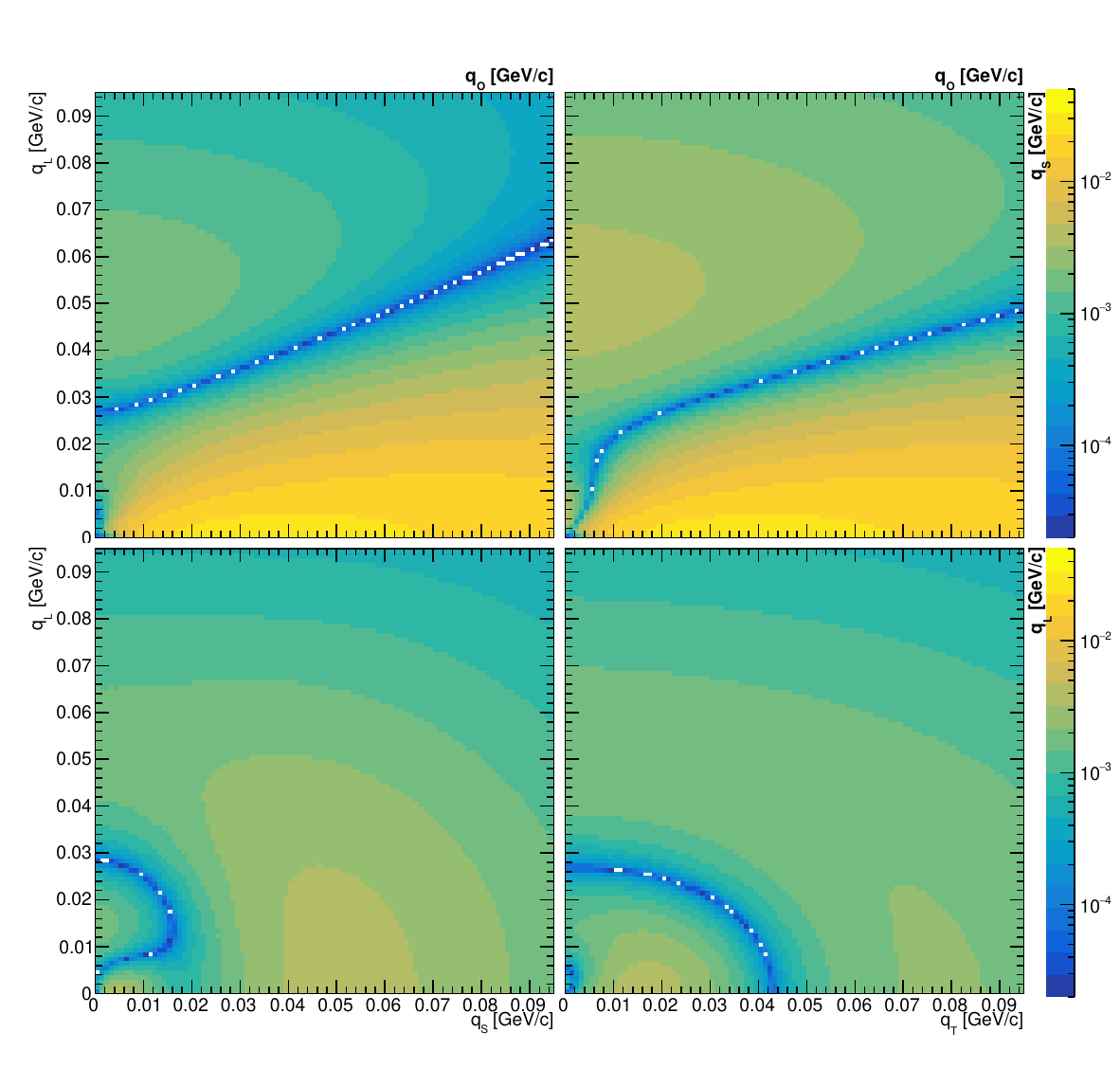}
\caption{Difference of the full and approximative 3D calculation in 2D slices, with the same parameters and $\beta_T$ as for Fig.~\ref{f:coulcorrcompare}. The four slices are in the planes $(q_{\rm out},q_{\rm long})$ (top left), $(q_{\rm out},q_{\rm side})$ (top right), $(q_{\rm side},q_{\rm long})$ (bottom left), $(q_{\rm T},q_{\rm long})$ (bottom right), with $q_{\rm T} = \sqrt{q_{\rm out}^2{+}q_{\rm side}^2}$}.
\label{f:coulcorrcompare.2D.Cdiff}
\end{figure*}

The above calculations show the importance of a full 3D calculation of the femtoscopic correlation functions when describing 3D correlation function measurements. It is furthermore important to note that such calculations are important also in case of a 1D measurement, when one employs the approximation of a spherically symmetric source. This approximation can for example be valid in heavy-ion collisions, but only in the LCMS. Thus when measuring 1D correlation functions, and describing them with calculations based on a spherically symmetric source, one often performs the measurement in the LCMS. However, the correlation function calculation is valid in the PCMS (i.e., the pair rest frame). Even if the above separation of the Coulomb correction and the pure quantum-statistical correlation function is utilized, in principle one should adapt the below formalism:
\begin{align}
&\quad C_2(Q;\lambda,R,\alpha) = 1{-}\lambda \,{+}\,\nonumber\\
&\quad +\lambda \cdot K_C(Q_{\rm PCMS}, R_{\rm PCMS},\alpha)\cdot C_2^{(0)}(Q;R,\alpha),
\label{e:C2Coulomblambda1D}
\end{align}
where~\cite{Kurgyis:2020vbz}
\begin{align}
\quad Q_{\rm PCMS} &\approx Q\sqrt{1\,{-}\,\frac{\beta_T^2}3},\\
\quad R_{\rm PCMS} &\approx R\sqrt{\frac{1\,{-}\,\frac23\beta_T^2}{1\,{-}\,\beta_T^2}}.
\end{align}
A more accurate result can, however, be achieved if even in this 1D setup one calculates the Coulomb correction in 3D, according to the boost to the PCMS:
\begin{align}
&\quad C_2(Q;\lambda,R,\alpha) = 1{-}\lambda \,{+}\,\nonumber\\
&\quad +\lambda\,{\cdot}\,\langle K_C({\bs Q}_{\rm PCMS},{\bs R}_{\rm PCMS},\alpha)\rangle\,{\cdot}\,C_2^{(0)}(Q;R,\alpha),
\label{e:C2Coulomblambda1DCoul3D}
\end{align}
where the $\langle \cdots \rangle$ around $K_C$ means that an angular average (w.r.t. the direction of ${\bs Q}_{\rm PCMS}$) is performed. It turns out, that in fact this angular averaging can be performed analytically, by substituting the spherically averaged version of $f(\bs Q)$, which can be denoted by $f_s(q)$ and can be calculated as
\begin{align}
\quad f_s(q) = \sint{-2pt}{-9pt}{-1}1\m dy\,\m{exp}\Bigg[{-}(q R_{\rm LCMS})^\alpha\bigg(1\,{+}\,\frac{\beta_T^2y^2}{1{-}\beta_T^2}\bigg)^{\hspace{-2pt}\frac\alpha2} \Bigg],
\label{e:fqsphsymm}
\end{align}
into the 1D variant of Eq.~\eqref{e:C2AA}. This spherically symmetric $f_s(q)$ is utilized in Eqs.~(40)-(41) of Ref.~\cite{Nagy:2023zbg}, and Eq.~(48) of the same paper gives the final result, similar to Eq.~\eqref{e:C2AA} of this paper.

\begin{figure}
\includegraphics[width=0.99\linewidth]{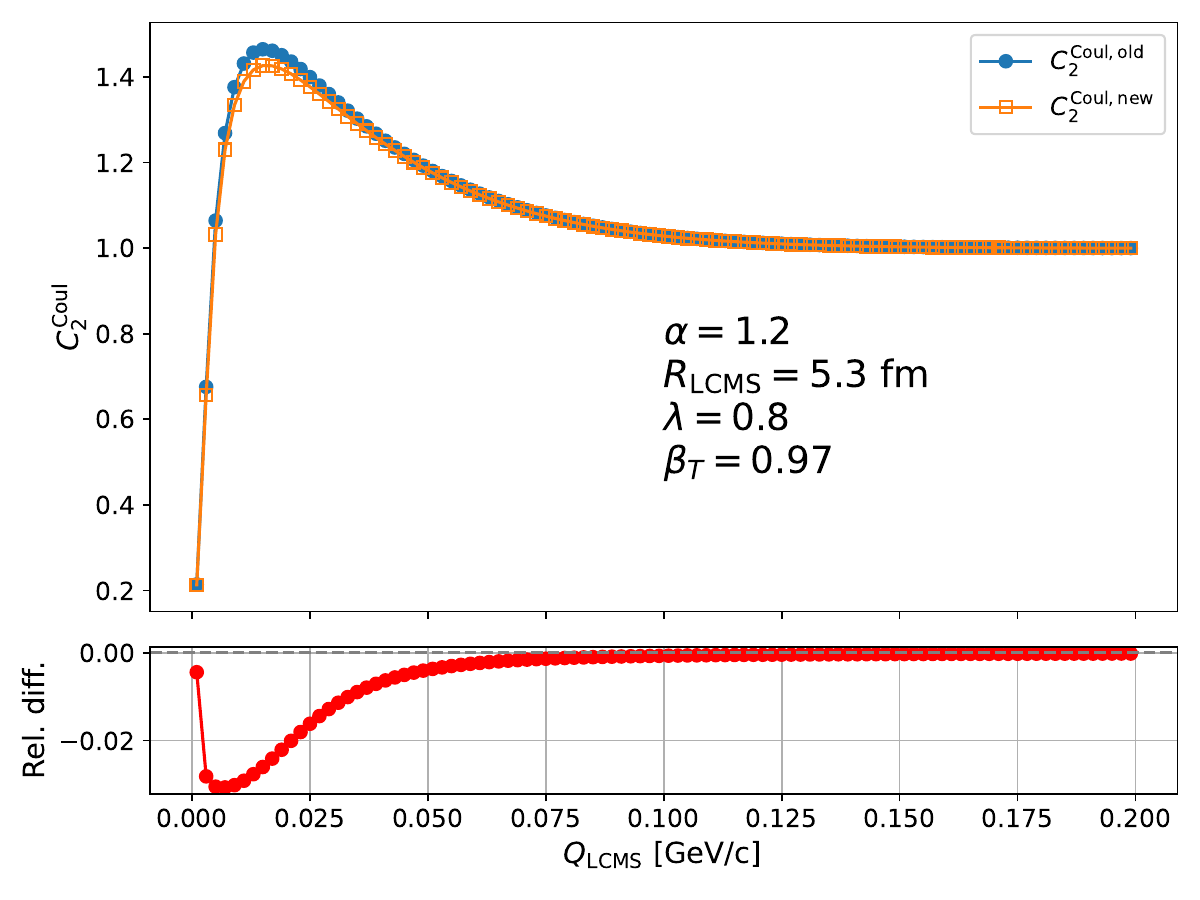}
\caption{Results of two versions of the 1D correlation function calculation: the usually utilized framework of Eq.~\eqref{e:C2Coulomblambda1D} with filled blue circles, and the enhanced approximation of this paper, given in Eq.~\eqref{e:C2Coulomblambda1DCoul3D}, with empty orange squares. Both calculations were performed at one specific $\beta_T$ value of 0.97, where the differences are more pronounced. The bottom panel shows the relative difference of the two curves as a function of $Q_{\rm LCMS}$.}
\label{f:betatest}
\end{figure}

\begin{figure}
\includegraphics[width=0.99\linewidth]{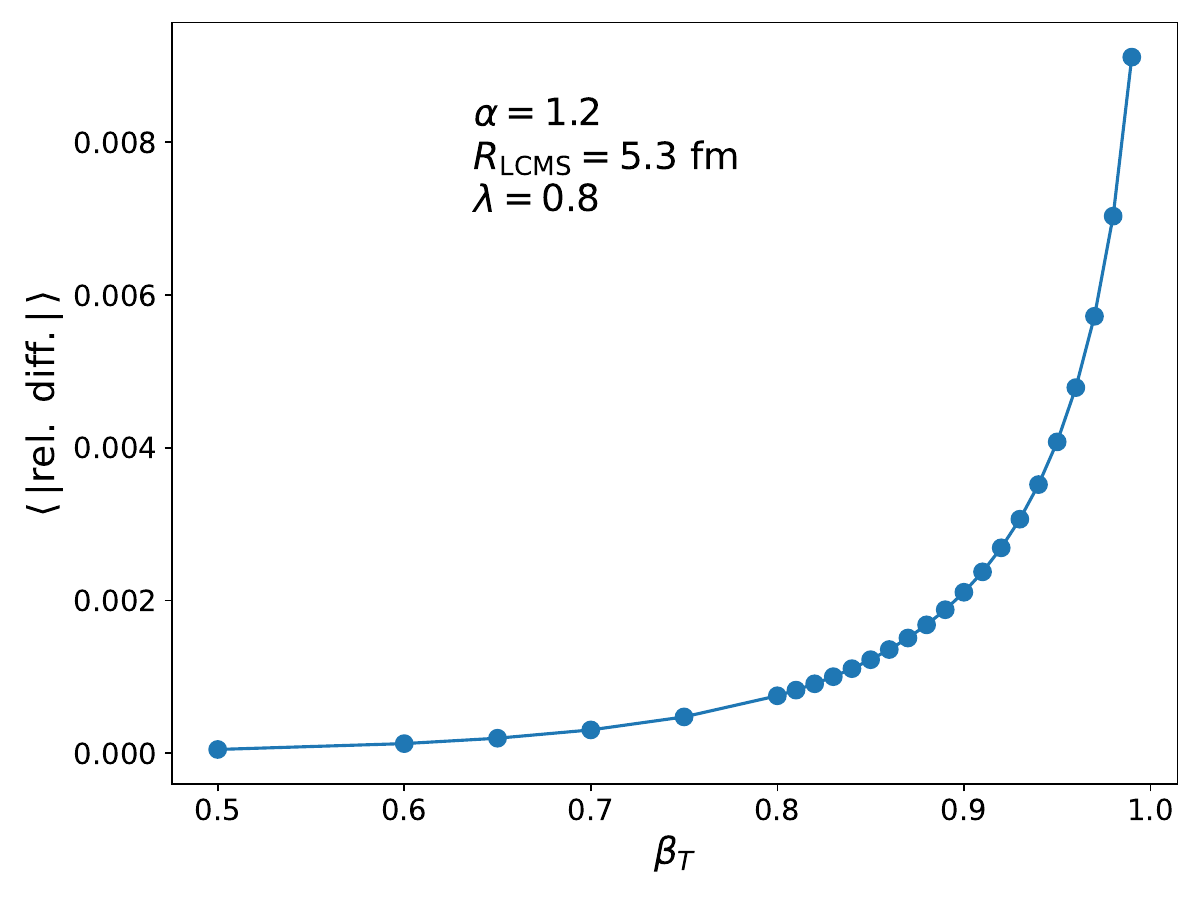}
\caption{The same comparision as Fig.~\ref{f:betatest}, but for various values of $\beta_T$ and averaged in the $Q_{\rm LCMS}<0.1$ GeV/$c$ range.}
\label{f:betatest.diff}
\end{figure}

The approximation utilizing the 1D $R_{\rm PCMS}$, Eq.~\eqref{e:C2Coulomblambda1D} is compared to the more accurate result that utilizes Eqs.~\eqref{e:C2Coulomblambda1DCoul3D} and \eqref{e:fqsphsymm}, in Fig.~\ref{f:betatest} for a given $\beta_T$, and then the average deviation as a function of $\beta_T$ is shown in Fig.~\ref{f:betatest.diff}. These results indicate that the simpler approximation is acceptable for moderate $\beta_T$ values. However, for larger momenta, where $\beta_T$ approaches one, the approximation of Eq.~\eqref{e:C2Coulomblambda1DCoul3D} can be used in an equally straigthforward but more accurate manner.

\section{Summary and outlook}\label{s:sum}

In this paper, we presented a self-consistent method to calculate Bose–Einstein correlation functions including the Coulomb final-state interaction for fully three-dimensional, non-spherical sources. We reformulated the Coulomb correction of Bose–Einstein correlation functions in Fourier space, allowing the correlation function to be expressed directly in terms of the source kernel without numerically unstable intermediate steps. The method is applicable to general, three-dimensional source distributions, with particular focus on elliptically contoured Lévy-stable sources motivated by recent experimental results. Explicit semi-analytic expressions were derived for the Coulomb correction terms using carefully chosen coordinate systems in momentum space. Utilizing the results, the accuracy of widely used spherical approximations was systematically tested against the full 3D calculation. It was found that these approximations work reasonably well for moderate transverse pair velocities but can lead to measurable deviations at large velocities or high anisotropies. Our study also clarified the role of frame transformations for Coulomb corrections. A ready-to-use numerical software package implementing the full 3D calculation is also provided. Overall, our work enables more accurate femtoscopic analyses for modern high-statistics heavy-ion data.

\begin{acknowledgements}
This research was funded by the NKFIH grants TKP2021-NKTA-64, PD-146589, K-146913, K-138136, and NKKP ADVANCED 152097.
\end{acknowledgements}

\FloatBarrier
\appendix

\section{Calculation of $\c D_{1\lambda}$ and $\c D_{2\lambda}$}\label{s:app:D12}

Here we derive the expressions of the $\c D_{1\lambda}$ and $\c D_{2\lambda}$, \Eqs{e:D1I1main}{e:D2I2main}, intermediate steps in our calculation of the correlation function in the body of the paper. The definitions in \Eqs{e:D1lambdadef}{e:D2lambdadef} imply that
\begin{align}
\label{e:D1I1}
\quad& \c D_{1\lambda}(\bs q) = -\td{}\lambda\c I_{1\lambda}(\bs q),\\
& \c D_{2\lambda}(\bs q) = -\td{}\lambda\c I_{2\lambda}(\bs q),
\label{e:D2I2}
\end{align}
where $\c I_{1\lambda}(\bs q)$ and $\c I_{2\lambda}(\bs q)$ are given by  
\begin{align}
&\quad \c I_{1\lambda}(\bs q) =\slint{-3pt}{2pt}{}\m d^3\bs r\,\frac{e^{-\lambda r}}r M\big(1{-}i\eta,1,i(kr{+}\bs k\bs r)\big)\times\nonumber\\
& \qquad\qquad\qquad\quad \times M\big(1{+}i\eta,1,-i(kr{+}\bs k\bs r)\big)e^{i\bs q\bs r},
\label{e:I1lambdadef}\\
&\quad \c I_{2\lambda}(\bs q) =\slint{-3pt}{2pt}{}\m d^3\bs r\,\frac{e^{-\lambda r}}r M\big(1{-}i\eta,1,i(kr{+}\bs k\bs r)\big)\times\nonumber\\
& \qquad\qquad\qquad\quad \times M\big(1{+}i\eta,1,-i(kr{-}\bs k\bs r)\big)e^{i\bs q\bs r}.
\label{e:I2lambdadef}
\end{align}
The reason for the insertion of this extra step of $-\td{}\lambda$ derivation is that the $\c I_{1\lambda}$, $\c I_{2\lambda}$ integrals are easier to calculate. Below we indeed do this; they turn out to be
\begin{align}
\label{e:I1result}
&\quad \c I_{1\lambda}(\bs q) = \frac{4\pi}{\lambda^2{+}q^2}\c U_{1-}\c F_+(x_1), \\
&\quad \c I_{2\lambda}(\bs q) = \frac{4\pi}{\lambda^2{+}q_+^2}\c U_{2-}\c F_+(x_2),
\label{e:I2result}
\end{align}
as already indicated in \Eqs{e:D1I1main}{e:D2I2main} in the main text. The definitions of $x_1$, $x_2$, $\c U_{1\pm}$, and $\c U_{2\pm}$ are found in Eqs.~\r{e:x1x2def}--\r{e:U2def}, and we use the hypergeometric functions $\c F_+$ and $\c F_-$ as defined in \Eqs{e:Fplusdef}{e:Fminusdef}. Later on we will also make some use of the hypergeometric functions with conjugate parameters,
\begin{align}
&\quad\overline{\c F}_+(x) := {}_2F_1({-}i\eta,1{-}i\eta,1,x),\\
&\quad\overline{\c F}_-(x):=(1{-}i\eta)\cdot{}_2F_1(i\eta,1{+}i\eta,2,x),
\end{align}
as well as relations that follow from known properties of the hypergeometric functions:
\begin{align}
&\quad\c F_+(x) = (1{-}x)^{-2i\eta}\overline{\c F}_+(x),\label{e:Fidentity1}\\
&\quad\c F_-(x) = (1{-}x)^{2i\eta}\big[2\c F_+(x)\,{-}\,\overline{\c F}_-(x)\big],\label{e:Fidentity2}\\
&\quad\c F'_+(x) = i\eta(1{-}x)^{-1-2i\eta}\c F_-(x);\label{e:Fderiv}
\end{align}
the last one being the reason for having introduced the $\c F_-$ parameter combination.

Performing the derivations w.r.t. $-\lambda$ in the expressions \r{e:I1result}--\r{e:I2result} of $\c I_{1\lambda}$ and $\c I_{2\lambda}$, using the expression for $\c F'_+$ from \r{e:Fderiv} above, and using the relations 
\begin{align}
&\quad \c U_{1-,2-} = (1{-}x_{1,2})^{2i\eta}\,\c U_{1+,2+},
\end{align}
after some simplification we have
\begin{align}
&\quad \c D_{1\lambda}(\bs q) = \frac{32i\pi\eta\,\c U_{1+}\c F_-(x_1)}{(\lambda^2{+}q^2)^2}
\times\nonumber\\
&\quad\qquad {\times}\frac{(\bs q\bs k{-}ik\lambda)[ik(q^2{-}\lambda^2){+}2\lambda\bs q\bs k]}{(\lambda^2{+}q^2)^2{+}4(\bs q\bs k{-}ik\lambda)^2} + \nonumber\\
&\qquad+\bigg[\frac{2\lambda(1{+}2i\eta)}{\lambda^2{+}q^2} -\frac{4i\eta(\lambda{-}ik) }{\lambda^2{+}\bs q\bs q_+{-}2ik\lambda}\bigg]\c I_{1\lambda}(\bs q),
\label{e:app:D1lambda}\\
&\quad \c D_{2\lambda}(\bs q) = \frac{32i\pi\eta\,\c U_{2+}\c F_-(x_2)}{(\lambda^2{+}q^2)(\lambda^2{+}q_+^2)}\times\nonumber\\
&\qquad\times\frac\lambda{\lambda^2{+}q_+^2}
\frac{[k^2q^2{-}(\bs q\bs k)^2](2\lambda^2{+}q^2{+}q_+^2)}{(\lambda^2{+}q^2)(\lambda^2{+}q_+^2){-}4[k^2q^2{-}(\bs q\bs k)^2]}+\nonumber\\
&{+}\bigg[\frac{2\lambda(1{+}i\eta)}{\lambda^2{+}q_+^2} {+}\frac{2\lambda i\eta}{\lambda^2{+}q^2}
{-} \frac{4i\eta(\lambda{-}ik)}{\lambda^2{+}\bs q\bs q_+{-}2ik\lambda}\bigg]\c I_{2\lambda}(\bs q).
\label{e:app:D2lambda}
\end{align}
For finite $\lambda{>}0$, these expressions are continuous functions of $\bs q$, and for $|\bs q|\,{\to}\,\infty$, decay at least as fast as $\sim\rec{q^4}$; to verify this, one needs to know that $|\c U_{1\pm}|$ and $|\c U_{2\pm}|$ are globally bounded by $e^{2\pi\eta}$, and that both $\c F_+(x)$ and $\c F_-(x)$ are also bounded functions (as can be verified using various known identities of the hypergeometric function). From these it follows that $\c D_{1\lambda}(\bs q)$ and $\c D_{2\lambda}(\bs q)$ are both integrable (over the whole $\bs q$-space), and as such, their integrals is given by the values of their respective Fourier transforms at the zero values of their respective variables. But as seen from \Eqs{e:D1lambdadef}{e:D2lambdadef}, $\c D_{1\lambda}(\bs q)$ and $\c D_{2\lambda}(\bs q)$ \textit{are} defined as Fourier transforms, and the integrands in those definitions are particularly simple to evaluate at $\bs r\,{=}\,0$. We thus have (inserting the $(2\pi)^3$ factor pertinent to the inverse properties of Fourier transforms)
\begin{align}
\quad \slint{-2pt}{2pt}{}\m d^3\bs q\,\c D_{1\lambda}(\bs q) = \slint{-2pt}{2pt}{}\m d^3\bs q\,\c D_{2\lambda}(\bs q) = 8\pi^3,
\label{e:Dlambdaint}
\end{align}
independent of $\lambda$, as long as $\lambda{>}0$.

We now turn to the calculation of $\c I_{1\lambda}$ and $\c I_{2\lambda}$, i.e. the integrals in \Eqs{e:I1lambdadef}{e:I2lambdadef}. The method is similar to that used by
Nordsieck~\cite{Nordsieck:1954zz} occuring in the theory of Bremsstrahlung, and was already recapitulated by us in Ref.~\cite{Nagy:2023zbg}, so here we content ourselves
with a somewhat streamlined exposition. One can use the following representation of confluent hypergeometric function:
\begin{align}
\quad M(a,1,z) = \rec{2\pi i}\soint{-2pt}{-10pt}{}{(0+,1+)}\m dt\,e^{tz}\rect t\big(1{-}\rect t\big)^{-a},
\end{align}
where the integration path on the complex $t$ plane encircles the branch cut on $t\,{\in}\,[0,1]$. Inserting this twice into the expressions of $\c I_{1\lambda}$ and
$\c I_{2\lambda}$ we have, paying attention to the order of integrals, 
\begin{align}
&\quad\c I_{1\lambda}(\bs q) = \sint{-3pt}{-2pt}{}{}\frac{\m d^3\bs r}r\soint{-2pt}{-3pt}{}{}\frac{\m du\,\m dt}{-4\pi^2}\,
e^{-\lambda r+i\bs q\bs r-i(t-u)(kr+\bs k\bs r)}    \times\nonumber\\&\qquad\qquad\qquad\times
\rect t\big(1{-}\rect t\big)^{-1-i\eta}\rect u\big(1{-}\rect u\big)^{-1+i\eta},\\
&\quad\c I_{2\lambda}(\bs q) = \sint{-3pt}{-2pt}{}{}\frac{\m d^3\bs r}r\soint{-2pt}{-3pt}{}{}\frac{\m du\,\m dt}{-4\pi^2}\,
e^{-\lambda r+i\bs q\bs r-i(t-u)kr+i(t+u)\bs k\bs r}\times\nonumber\\&\qquad\qquad\qquad\times
\rect t\big(1{-}\rect t\big)^{-1-i\eta}\rect u\big(1{-}\rect u\big)^{-1+i\eta},
\end{align}
Investigating the repeated integrability it turns out that if additional restrictions are placed on the paths of the complex $t$ and $u$ variables, we can interchange
the integrals, and perform the integral over $\bs r$ first:
\begin{align}
\label{e:I1paths}
&\quad\c I_{1\lambda}(\bs q) = \soint{-2pt}{-3pt}{}{}\frac{\m du\,\m dt}{-4\pi^2ut}\big(1{-}\rect t\big)^{-1-i\eta}\big(1{-}\rect u\big)^{-1+i\eta}\times\nonumber\\
&\qquad\qquad\quad\times\sint{-3pt}{-2pt}{}{}\frac{\m d^3\bs r}re^{-\lambda r+i\bs q\bs r-i(t-u)(kr+\bs k\bs r)}\nonumber\\
&\quad\qquad\qquad\qquad\tn{if}\;\;\m{Im}(t{-}u)\,{<}\,\fract\lambda{2k},\\ 
&\quad\c I_{2\lambda}(\bs q) = \soint{-2pt}{-3pt}{}{}\frac{\m du\,\m dt}{-4\pi^2ut}\big(1{-}\rect t\big)^{-1-i\eta}\big(1{-}\rect u\big)^{-1+i\eta}\times\nonumber\\
&\qquad\qquad\quad\times\sint{-3pt}{-2pt}{}{}\frac{\m d^3\bs r}re^{-\lambda r+i\bs q\bs r-i(t-u)kr+i(t+u)\bs k\bs r)}\nonumber\\
&\quad\qquad\qquad\qquad\tn{if}\;\;\m{Im}\,t\,{<}\,\fract\lambda{2k}\;\,\tn{and}\;\,\m{Im}\,u \,{>}\,{-}\fract\lambda{2k}.
\label{e:I2paths}
\end{align}
These conditions are to be fulfilled here by any $u$ and $t$ encountered on their respective integration paths. Now the $\bs r$-integral can be performed with the formula
\begin{align}
\quad\sint{0pt}{-2pt}{}{}\frac{\m d^3\bs r}re^{-\beta r+i\bs B\bs r}=\frac{4\pi}{\beta^2{+}\bs B^2},
\label{e:spaceint1}
\end{align}
valid for any $\beta\,{\in}\,\mathbb C$ and any $\bs B\,{\equiv}\,(B_x,B_y,B_z)$ with any complex components, if $\m{Re}\,\beta \,{>}\,|\m{Im}\,\bs B|$.%
\footnote{
Here (as it is clear on grounds of analyticity in $\beta$, $B_x$, $B_y$ and $B_z$) the meaning of $\bs B^2$ is to be understood as $B_x^2{+}B_y^2{+}B_z^2$ even for complex
values of the components (in particular, no conjugation is needed). Also, $\m{Im}\,\bs B$ is understood as $(\m{Im}\,B_x,\m{Im}\,B_y,\m{Im}\,B_z)$, and $|\m{Im}\,\bs B|$
is its length as that of a standard real three-vector.
}
With this done, we have (with the same conditions on $u$ and $t$)
\begin{align}
&\quad\c I_{1\lambda}(\bs q) = -\rec\pi\soint{-2pt}{-2pt}{}{}\frac{\m du\,\m dt}{ut}\big(1{-}\rect t\big)^{-1-i\eta}\big(1{-}\rect u\big)^{-1+i\eta}\times\nonumber\\
&\qquad\times\rec{\lambda^2{+}q^2{-}2(\bs q\bs k{-}ik\lambda)(t{-}u)} \label{e:I1},\\ 
&\quad\c I_{2\lambda}(\bs q) = -\rec\pi\soint{-2pt}{-2pt}{}{}\frac{\m du\,\m dt}{ut}\big(1{-}\rect t\big)^{-1-i\eta}\big(1{-}\rect u\big)^{-1+i\eta}\times\nonumber\\
&\qquad\times\rec{\lambda^2{+}q^2{+}2\bs q\bs k(t{+}u){+}2ik\lambda(t{-}u){+}4utk^2}. \label{e:I2}
\end{align}
In both cases, we can now perform the $u$-integral at a fixed $t$ by expanding the $u$-contour to infinity, since the $u$-integrand decreases as $\sim\rec{u^2}$. As a
function of $u$, both integrands have, apart from the encircled branch cut on $u\,{\in}\,[0,1]$, one simple pole at $u_{1t}$ and $u_{2t}$, respectively:  
\begin{align}
&\quad u_{1t} = t-\rec2\frac{\lambda^2{+}q^2}{\bs q\bs k{-}ik\lambda},\\
&\quad u_{2t} = -\rec2\frac{q^2{+}2\bs q\bs k\big(t{-}\frac{i\lambda}{2k}\big)}{\bs q\bs k{+}2k^2t{-}ik\lambda}-\frac{i\lambda}{2k},
\end{align}
and from the following easily verifiable relations, 
\begin{align}
\label{e:u1tconstraint}
&\quad\m{Im}\big[t{-}u_{1t}\big]\,{-}\,\frac\lambda{2k} = \frac\lambda{2k}\bigg|\frac{\bs q{\times}\bs k}{\bs q\bs k{-}ik\lambda}\bigg|^2,\\
&\quad\m{Im}\,u_{2t}\,{+}\,\frac\lambda{2k}
= \bigg|\frac{\bs q{\times}\bs k}{\bs q\bs k{+}2k^2t{-}ik\lambda}\bigg|^2\hspace{-2pt}{\cdot}\,\bigg[\m{Im}\,t\,{-}\,\frac\lambda{2k}\bigg],
\label{e:u2tconstraint}
\end{align}
one concludes that both $u_{1t}$ and $u_{2t}$ lie on the $u$ plane outside of the respective $u$-integration paths that were up to now, for the sake of the interchangeability of the integrals, constrained by the conditions stated in \Eqs{e:I1paths}{e:I2paths}.%
\footnote{
The moral of the relations in \Eqs{e:u1tconstraint}{e:u2tconstraint} is that for any $t$ satisfying the conditions on the $t$-path, $u_{1t}$ and $u_{2t}$ cannot satisfy the requirement on the $u$-path, so no allowable $u$-path can cross $u_{1t}$ and $u_{2t}$. But if they were inside a given $u$-path, tightening this path around the $[0,1]$ branch cut (allowable by any means) would make the paths cross $u_{1t}$ and $u_{2t}$.
}
So by expanding the $u$-paths to infinity, the $u$-integrals yield $-2\pi i$ times the residues at $u_{1t}$ and $u_{2t}$, respectively. We thus have
\begin{align}
&\quad\c I_{1\lambda}(\bs q) = i\soint{-1pt}{-2pt}{}{}\frac{\m dt}t\frac{\big(1{-}\rect t\big)^{-1-i\eta}}{\bs q\bs k{-}ik\lambda}
\frac{\big(1{-}\rect{u_{1t}}\big)^{i\eta}}{u_{1t}\,{-}\,1},\\
&\quad\c I_{2\lambda}(\bs q) = i\soint{-1pt}{-2pt}{}{}\frac{\m dt}t\frac{\big(1{-}\rect t\big)^{-1-i\eta}}{\bs q\bs k{-}ik\lambda{+}2k^2t}
\frac{\big(1{-}\rect{u_{2t}}\big)^{i\eta}}{u_{2t}\,{-}\,1},
\end{align}
with the conditions in \r{e:I1paths}--\r{e:I2paths} on the respective $t$-paths still in force: these mean in effect that $\m{Im}\,t\,{<}\,\frac\lambda{2k}$ must hold in
both cases. These last integrands as a function of $t$ have two branch cuts; in both cases one, the original $[0,1]$ line segment, is encircled by the integration path,
while the other one, a circle segment connecting the $u_{1,2t}{=}0$ and $u_{1,2t}{=}1$ points, lies outside of the contour (as can be verified by inequalities similar to
\Eqs{e:u1tconstraint}{e:u2tconstraint}. These last integrals can be cast into forms whose integrands contain three singular points (to discover a representation of the
ordinary hypergeometric function) by a substitution $t\,{=}\,\rec s$, with which we get similar forms in the two cases:
\begin{align*}
\quad\c I_{1,2\lambda}(\bs q) \,{=}\, \frac{-2i}{\lambda^2{+}q^2}\soint{-3pt}{-2pt}{}{}\m ds\frac{(1{-}s)^{-1-i\eta}}{\beta(s{-}B_{1,2})}
\bigg[\beta\frac{s{-}B_{1,2}}{s{-}A_{1,2}}\bigg]^{i\eta},
\end{align*}
where the notations are 
\begin{align}
&\quad \beta=1{+}2\frac{\bs q\bs k{-}ik\lambda}{\lambda^2{+}q^2},\\
&\quad A_1=2\frac{\bs q\bs k{-}ik\lambda}{\lambda^2{+}q^2},  \quad\;\;\, B_1=2\frac{\bs q\bs k{-}ik\lambda}{\lambda^2{+}\bs q\bs q_+{-}2ik\lambda},\\
&\quad A_2=-2\frac{\bs q\bs k{+}ik\lambda}{\lambda^2{+}q^2}, \;\;\;      B_2=-2\frac{\bs q\bs k{+}ik\lambda{+}2k^2}{\lambda^2{+}\bs q\bs q_+{-}2ik\lambda},
\end{align}
and the integration path on the $s$ plane does not intersect the $s\,{\in}\,[1,\infty]$ half line but encircles $s{=}0$ and the branch cut between $s\,{=}\,A_{1,2}$ and
$s\,{=}\,B_{1,2}$. The next step is a linear $s\,{\to}\,\tau$, $s\,{=}\,B_{1,2}\,{+}\,(1{-}B_{1,2})\tau$ transformation that brings the three singular points $s{=}1$,
$s{=}B_{1,2}$ and $s{=}A_{1,2}$ into $\tau{=}1$, $\tau{=}0$ and $\tau{=}x_{1,2}$, where $x_1$ and $x_2$ are the variables defined in \Eq{e:x1x2def}, to arrive at
\begin{align*}
\quad\c I_{1,2\lambda} = 2\sointl{-10pt}{-10pt}{}{(0+,x+)}\m d\tau\,\frac{((1{-}B_{1,2})(1{-}\tau))^{-1-i\eta}}{i(\lambda^2{+}q^2)\beta\tau}
\bigg(\frac{\beta\tau}{\tau{-}x_{1,2}}\bigg)^{i\eta}.
\end{align*}
The last step is to carefully investigate the phases of the factors to ensure that identities like $(ab)^c = a^cb^c$ can be used to simplify the
integrand.%
\footnote{
For complex $a$, $b$, $c$, $(ab)^c$ is not necessarily equal to $a^cb^c$, since the phase of $ab$ may not be equal to the sum of that of $a$ and $b$ because of the possible
$\pm2\pi$ shift when traversing the negative real line with a multiplication, to ensure that all phases customarily fall into the $]{-}\pi,\pi]$ interval. 
}
In doing so, at an intermediary step it is necessary to transform the integration path on the $\tau$ plane from a closed one to one that goes from $-\infty$ back to
$-\infty$ while going around $\tau{=}0$ and $\tau{=}x_{1,2}$. The final form is thus
\begin{align}
&\quad \c I_{1,2\lambda} = \frac{-2i(1{-}B_{1,2})^{-1-i\eta}}{(\lambda^2{+}q^2)\beta^{1-i\eta}}\times\nonumber\\
&\qquad\times\sint{-2pt}{-24pt}{\tau=-\infty}{(0+,x+)}\m d\tau\,\tau^{-1+i\eta}(1{-}\tau)^{-1-i\eta}(\tau{-}x_{1,2})^{-i\eta}.
\label{e:I12last}
\end{align}
The goal of the manipulations with the single remaining $t$-integral was to be able to use the following representation of the ordinary hypergeometric function: 
\begin{align}
\label{e:hypintrepresent}
&\quad{}_2F_1\big(a,b,c,z\big) = \frac{\Gamma(c)\Gamma(b{-}c{+}1)}{2\pi i\cdot\Gamma(b)}\times\nonumber\\
&\qquad\qquad\times\sint{-2pt}{-24pt}{\tau=-\infty}{(0+,z+)}\m d\tau\,\tau^{a-c}(1{-}\tau)^{c-b-1}(\tau{-}z)^{-a}.
\end{align}
With this from \Eq{e:I12last} we arrive at the results for $\c I_{1\lambda}$ and $\c I_{2\lambda}$ as in \Eqs{e:I1result}{e:I2result} that lead to \Eqs{e:D1I1main}{e:D2I2main} for $\c D_{1\lambda}$ and $\c D_{2\lambda}$, or in a more explicit form, in \Eqs{e:app:D1lambda}{e:app:D2lambda}.

\section{The $\lambda{\to}0$ limit of $\c D_{1\lambda}$ and $\c D_{2\lambda}$}\label{s:app:D12limit}

The toolkit to perform the $\lambda\,{\to}\,0$ limit in \Eq{e:CQD} is again \textit{Lebesgue's theorem}, recited here in the body of the paper after \Eq{e:CDf1}. The goal is to transform the integrals into a form where one can find a dominant function called for by the theorem, so that the limit can be performed under the integration sign, leaving only the integral (and not the limit) finally. In the following we write up estimations (inequalities) about the moduli of integrands looking for such dominants (valid for any $\lambda$); whenever the dominant is integrable, and the original integrand converges to 0 almost everywhere pointwise, then the limit of integral is 0. If we apply this to a difference of two integrands, then we can substitute one to the another (presumably a more compicated one to a simpler one).

As a first step, we separate the values of $f(\bs q)$ at $\bs q\,{=}\,\bs 0$ and at $\bs q\,{=}\,{-}\bs Q$ in the terms in \Eq{e:CQD} containing $\c D_{1\lambda}$ and $\c D_{2\lambda}$, respectively. Retrospectively, this might be suggested by the fact that these values of $f$ play such emphasized role in the result for the corresponding terms in the interaction-free ($\eta{=}0$) case, as seen from Eqs.~\r{e:Deta0}--\r{e:Cfreecheck}. As seen below, this subtraction turns out to be a crucial idea towards the solution, as it enables us to perform the limits in the integrals; otherwise the estimations needed for the application of the Lebesgue theorem indeed would not have worked. We thus proceed from \Eq{e:CQD} as
\begin{align}
\quad C(\bs Q)\,{=}\,\frac{|\c N|^2}{8\pi^3}\lim_{\lambda\to0}\slint{-1pt}{1pt}{}\m d^3\bs q\,\big[f(\bs q){-}f(\bs 0)\big]\c D_{1\lambda}(\bs q) +\nonumber\\
\qquad +\frac{|\c N|^2}{8\pi^3}\lim_{\lambda\to0}\slint{-1pt}{1pt}{}\m d^3\bs q\,\big[f(\bs q){-}f(-\bs Q)\big]\c D_{2\lambda}(\bs q) +\nonumber\\
+\frac{|\c N|^2}{8\pi^3}\hspace{-2pt}\lim_{\lambda\to0}\slint{-1pt}{1pt}{}\m d^3\bs q\Big[f(\bs 0)\c D_{1\lambda}(\bs q)\,{+}\,f(-\bs Q)\c D_{2\lambda}(\bs q)\Big].
\end{align}
In the last terms, the integrals of $\c D_{1\lambda}$ and $\c D_{2\lambda}$ are evaluated using \Eq{e:Dlambdaint} above; that result is valid for any $\lambda\,{>}\,0$, so clearly it is in the $\lambda\,{\to}\,0$ limit as well. We thus arrive at a form displayed in \Eq{e:C2AA} as
\begin{align}
\quad C(\bs Q) = |\c N|^2\Big[1 \,{+}\, f(-\bs Q) \,{+}\, \frac\eta\pi\big[\c A_1\,{+}\,\c A_2\big]\Big],
\end{align}
where the $\c A_1$ and $\c A_2$ quantities are
\begin{align}
\label{e:A1def0}
\quad \c A_1=\rec{8\pi^2\eta}\lim_{\lambda\to0}\slint{-2pt}{2pt}{}\m d^3\bs q\,\c D_{1\lambda}(\bs q)f_1(\bs q),\\
\quad \c A_2=\rec{8\pi^2\eta}\lim_{\lambda\to0}\slint{-2pt}{2pt}{}\m d^3\bs q\,\c D_{2\lambda}(\bs q)f_2(\bs q),
\label{e:A2def0}
\end{align}
and we use the ,,reduced'' versions of the $f$ function as
\begin{align}
&\quad f_1(\bs q) = f(\bs q)\,{-}\,f(\bs 0),\\
&\quad f_2(\bs q) = f(\bs q)\,{-}\,f(-\bs Q).
\end{align}
We distributed the $\eta$ and $\pi$ factors in the definitions of $\c A_1$ and $\c A_2$ to match the corresponding formulas for the spherically symmetric case explored in Ref.~\cite{Nagy:2023zbg}. For the $\lambda$ parameter, we fix a ,,starting'' $\lambda_0$ value; we can and do require (for the simplicity of some later estimations) that $0\,{<}\,\lambda\,{<}\,\lambda_0\,{<}\,2k$ holds (with strict inequalities).

Below we state (but do not elaborate on) the integrability of functions deemed to be appropriate dominants. Integrability usually hinges on fast enough decrease at infinity, and on ,,spikes'' around specific points not converging to infinity too fast. For example, in one dimension, $|x|^{-\beta}$ is integrable around $|x|\,{\to}\,\infty$ if and only if $\beta\,{>}\,1$, while $|x{-}x_0|^{-\beta}$ is integrable around $x_0$ if and only if $\beta\,{<}\,1$ (strictly smaller). In more dimensions one has to take care of the Jacobi determinant: for example, if $\bs q$ is a three-dimensional variable, $|\bs q|^{-\beta}$ is integrable around $\bs q\,{\to}\,\infty$ if and only if $\beta\,{>}\,3$, and $|\bs q{-}\bs q_0|^{-\beta}$ is integrable around $\bs q_0$ if and only if $\beta\,{<}\,3$.

In our case, the behavior at infinity is usually not problematic: for example, polynomial and rational fractional functions of $\bs q$ can be overestimated by the leading order term (times a constant) at large enough $q$. We use the $\c K$ notation below for such constants and the ,,$q{>}Q$'' condition (with a fix $Q$ value safely above $2k$, $\lambda_0$, and $\lambda$; for example $Q\,{=}\,4k$) to denote such estimations (clearly valid for any allowed $\lambda$). On the other hand, neighbourhoods of particular points or surfaces may need particular attention.

We already remarked that $\c U_{1\pm}$ and $\c U_{2\pm}$ as well as $\c F_\pm(x_1)$ and $\c F_\pm(x_2)$ are globally bounded; i.e. there exists a $\c C$ constant with which for any $\lambda$ and any $\bs q$, 
\begin{align}
\label{e:UFestim}
\quad |\c U|\le\c C, \quad |\c F|\le\c C, \quad |\c U\c F| \le \c C,
\end{align}
for any combinations of indices and $\pm$ signs. We frequently use the following inequality, valid for any $x\,{\in}\,\B R$:
\begin{align}
\quad \frac\lambda{\lambda^2{+}x^2}\le\m{min}\bigg\{\frac{\lambda_0}{x^2},\rec{2|x|}\bigg\};
\label{e:lorentzestim}
\end{align}
or more generally, for any $0\,{<}\,\xi\,{<}\,1$:
\begin{align}
\quad \frac{\lambda^{1+\xi}}{\lambda^2{+}x^2}\le\m{min}\bigg\{\frac{\lambda_0^{1+\xi}}{x^2},\rec{2^{1+\xi}|x|^{1-\xi}}\bigg\}.
\label{e:lorentzestim2}
\end{align}

We assumed $f(\bs q)$ to be bounded, integrable and continuous; now (just as in Ref.~\cite{Nagy:2023zbg}) we place an additional technical restriction.%
\footnote{
This is not necessarily the most general condition which would allow one to proceed with the derivation, although a fairly general one. It is fulfilled by any $f$ that is continuously differentiable, and also by the one that corresponds to L\'evy distributions.
}
We require that it has ,,at most power-law-type spikes''; i.e. there exists a $\sigma\,{>}\,0$ exponent and $K$ constant so that for any $\bs q$ and $\tilde{\bs q}$,
\begin{align}
\quad |f(\bs q){-}f(\tilde{\bs q})|\le K|\bs q{-}\tilde{\bs q}|^\sigma.
\label{e:estim:fqq}
\end{align}
We can combine this with $f$ being bounded to arrive at
\begin{align}
\label{e:estim:f2K1}
&\quad \big|f_1(\bs q)\big|\le\m{min}\big\{Kq^\sigma,  K'\big\},\\
&\quad \big|f_2(\bs q)\big|\le\m{min}\big\{Kq_+^\sigma,K'\big\},\\
&\quad \big|f_1(\bs q){-}f_1(\tilde{\bs q})\big|\le\m{min}\big\{K|\bs q{-}\tilde{\bs q}|^\sigma,K'\big\},\\
&\quad \big|f_2(\bs q){-}f_2(\tilde{\bs q})\big|\le\m{min}\big\{K|\bs q{-}\tilde{\bs q}|^\sigma,K'\big\},
\label{e:estim:f2K2}
\end{align}
with appropriate $K$, $K'$ constants; from these (and the functional form of real fractional power functions) it follows that if indeed there exists such $\sigma\,{>}\,0$ exponent, then we can take this to be as close to 0 as necessary (and choose the $K$ constant accordingly).

Substituting $\c D_{1\lambda}$ and $\c D_{2\lambda}$ from \Eqs{e:app:D1lambda}{e:app:D2lambda} into \Eqs{e:A1def0}{e:A2def0} we have
\begin{align}
&\quad\c A_1 = \rec\pi\lim_{\lambda\to0}\slint{-1pt}{1pt}{}\m d^3\bs q\frac{f_1(\bs q)}{\lambda^2{+}q^2}\Bigg\{
\frac{\lambda(1{+}2i\eta)}{\eta(\lambda^2{+}q^2)}\c U_{1-}\c F_+(x_1) {-}\nonumber\\
&-2\frac{(i\lambda{+}k)\c U_{1-}\c F_+(x_1)}{\lambda^2{+}\bs q\bs q_+{-}2ik\lambda} +\c U_{1+}\c F_-(x_1)\;\cdot\nonumber\\
&{\cdot}\,\bigg[\frac{i\lambda{-}k}{\lambda^2{+}\bs q\bs q_-{+}2ik\lambda} {+} \frac{i\lambda{+}k}{\lambda^2{+}\bs q\bs q_+{-}2ik\lambda}
{-}\frac{2i\lambda}{\lambda^2{+}q^2}\bigg]\Bigg\},
\label{e:R1defb}\\
&\quad\c A_2 = \rec\pi\lim_{\lambda\to0}\slint{-1pt}{1pt}{}\m d^3\bs q\frac{f_2(\bs q)}{\lambda^2{+}q_+^2}\Bigg\{
\frac{\lambda(1{+}i\eta)}{\eta(\lambda^2{+}q_+^2)}\c U_{2-}\c F_+(x_2) + \nonumber\\
&+\frac{i\lambda\,\c U_{2-}\c F_+(x_2)}{\lambda^2{+}q^2} -
\frac{2(i\lambda{+}k)\c U_{2-}\c F_+(x_2)}{\lambda^2{+}\bs q\bs q_+{-}2ik\lambda} +\c U_{2+}\c F_-(x_2)\,\cdot\nonumber\\
&\cdot\frac{2i\lambda[q^2q_+^2{-}(\bs q\bs q_+)^2](\lambda^2{+}\bs q\bs q_+{+}2k^2)}
{(\lambda^2{+}q^2)(\lambda^2{+}q_+^2)\big[(\lambda^2{+}\bs q\bs q_+)^2{+}4k^2\lambda^2\big]}\Bigg\}.
\label{e:R2defb}
\end{align}
All the terms in the integrands that contain a $\lambda$ factor in the numerator, except for the last term of $\c A_2$, yield zero for $\lambda\,{\to}\,0$. This follows from them converging pointwise to 0 everywhere except for zero-measure sets (the points $\bs q\,{=}\,0$ and/or $\bs q_\pm\,{=}\,0$ or the spherical surfaces $\bs q\bs q_\pm\,{=}\,0$), and that (as seen from Eqs.~\r{e:UFestim}, \r{e:lorentzestim}, \r{e:estim:f2K1}--\r{e:estim:f2K2}, and other simple estimations) they are dominated by $\lambda$-independent integrable functions (the right hand sides of the following inequalities):
\begin{align}
&\quad \bigg|\frac{\lambda\,\c U_{1\pm}\c F_\mp(x_1)}{(\lambda^2{+}q^2)^2}f_1(\bs q)\bigg|
\,{\le}\,\m{min}\bigg\{\frac{\c CK\lambda_0}{q^{4-\sigma}},\frac{\c CK}{2q^{3-\sigma}}\bigg\},\\ 
&\quad \bigg|\frac{\lambda\,\c U_{2-}\c F_+(x_2)}{(\lambda^2{+}q_+^2)^2}f_2(\bs q)\bigg|
\,{\le}\,\m{min}\bigg\{\frac{\c CK\lambda_0}{q_+^{4-\sigma}},\frac{\c CK}{2q_+^{3-\sigma}}\bigg\},\\ 
&\quad\bigg|\frac{\lambda\,\c U_{2-}\c F_+(x_2)\,f_2(\bs q)}{(\lambda^2{+}q^2)(\lambda^2{+}q_+^2)}\bigg|\le\frac{\lambda_0\c CK}{q^{2-\sigma}q_+^2},\\
&\quad \bigg|\fracd{f_1(\bs q)}{\lambda^2{+}q^2}\fracd{\lambda\,\c U_{1\pm}\c F_\mp(x_1)}{\lambda^2{+}\bs q\bs q_+{-}2ik\lambda}\bigg|\,{\le}\,
\Bigg\{\hspace{-3pt}
\begin{array}{l}
\frac{\c CK}{q^{2-\sigma}}\frac{\c K\lambda_0}{q^2}\;\tn{if $q{>}Q$,}\vspace{1pt}\\
\frac{\c CK}{2kq^{2-\sigma}}\;\tn{always,}
\end{array}\\
&\quad \bigg|\fracd{f_1(\bs q)}{\lambda^2{+}q^2}\fracd{\lambda\,\c U_{1+}\c F_-(x_1)}{\lambda^2{+}\bs q\bs q_-{+}2ik\lambda}\bigg|\,{\le}\, 
\Bigg\{\hspace{-3pt}
\begin{array}{l}
\frac{\c CK}{q^{2-\sigma}}\frac{\c K\lambda_0}{q^2}\;\tn{if $q{>}Q$,}\vspace{1pt}\\
\frac{\c CK}{2kq^{2-\sigma}}\;\tn{always,}
\end{array}\\
&\quad \bigg|\fracd{f_2(\bs q)}{\lambda^2{+}q_+^2}\fracd{\lambda\,\c U_{2-}\c F_+(x_1)}{\lambda^2{+}\bs q\bs q_+{-}2ik\lambda}\bigg|\,{\le}
\Bigg\{\hspace{-3pt}
\begin{array}{l}
\frac{\c CK}{q_+^{2-\sigma}}\frac{\c K\lambda_0}{q^2}\;\tn{if $q{>}Q$,}\vspace{1pt}\\
\frac{\c CK}{2kq_+^{2-\sigma}}\;\tn{always.}
\end{array}
\end{align}
In a similar way, we can leave the $\lambda^2{+}\bs q\bs q_+$ term in the numerator of the last fraction in \Eq{e:R2defb}, since (now leaving the universally bounded $\c U$ and $\c F(x)$ factors)
\begin{align}
&\quad\bigg|\frac{[q^2q_+^2{-}(\bs q\bs q_+)^2]}{(\lambda^2{+}q^2)(\lambda^2{+}q_+^2)^2}\frac{f_2(\bs q)\,\lambda(\lambda^2{+}\bs q\bs q_+)}
{(\lambda^2{+}\bs q\bs q_+)^2{+}4k^2\lambda^2}\bigg|\le\frac{2q^2q_+^2}{q^2q_+^4}\times \nonumber\\&\qquad\times
\frac{K'\,\lambda|\lambda^2{+}\bs q\bs q_+|}{(\lambda^2{+}\bs q\bs q_+)^2{+}4k^2\lambda^2}
\,{\le}\Bigg\{\hspace{-3pt}\begin{array}{l}
\frac{K'}{q_+^2}\frac{\c K\lambda_0}{q^2}\;\tn{if $q{>}Q$,}\vspace{1pt}\\
\frac2{q_+^2}\frac{K'}{2k}\;\;\tn{always.}
\end{array}
\end{align}
We thus have with the simplifications listed so far
\begin{align}
&\quad \c A_1 = \frac k\pi\lim_{\lambda\to0}\slint{-1pt}{1pt}{}\m d^3\bs q\frac{f_1(\bs q)}{\lambda^2{+}q^2}
\bigg\{\frac{-\c U_{1+}\c F_-(x_1)}{\lambda^2{+}\bs q\bs q_-{+}2ik\lambda} +\nonumber\\ 
&\qquad + \frac{\c U_{1+}\c F_-(x_1)\,{-}\,2\c U_{1-}\c F_+(x_1)}{\lambda^2{+}\bs q\bs q_+{-}2ik\lambda}\bigg\},
\label{e:A1def}\\
&\quad \c A_2 = \frac k\pi\lim_{\lambda\to0}\slint{-1pt}{1pt}{}\m d^3\bs q
\frac{f_2(\bs q)}{\lambda^2{+}q_+^2}\bigg\{\frac{-2\c U_{2-}\c F_+(x_2)}{\lambda^2{+}\bs q\bs q_+{-}2ik\lambda} + \nonumber\\ &\qquad 
+\frac{4i\lambda k\,\c U_{2+}\c F_-(x_2)}{(\lambda^2{+}\bs q\bs q_+)^2{+}4k^2\lambda^2}\frac{q^2q_+^2{-}(\bs q\bs q_+)^2}{(\lambda^2{+}q^2)(\lambda^2{+}q_+^2)}\bigg\}.
\label{e:A2def}
\end{align}
The last fraction in this expression of $\c A_2$ can be replaced by 1: the difference of the integrands converges to 0 almost everywhere (i.e. if $\bs q\bs q_+{\neq}\,0$), and is dominated by an integrable function:
\begin{align}
&\quad\bigg|\frac{f_2(\bs q)}{\lambda^2{+}q_+^2}\frac{\lambda \,\c U_{2+}\c F_-(x_2)}{(\lambda^2{+}\bs q\bs q_+)^2{+}4k^2\lambda^2}
\bigg[1{-}\frac{q^2q_+^2{-}(\bs q\bs q_+)^2}{[\lambda^2{+}q^2][\lambda^2{+}q_+^2]}\bigg]\bigg| =\nonumber\\ &\quad=
\bigg|\frac{\lambda \,f_2(\bs q)\,\c U_{2+}\c F_-(x_2)}{(\lambda^2{+}q^2)(\lambda^2{+}q_+^2)^2}\bigg|
\le \Bigg\{\hspace{-3pt}
\begin{array}{l}
\frac{\c CK'\c K\lambda_0}{q^6}\;\tn{if $q{>}Q$,}\vspace{1pt}\\
\frac{\c CK}{2q_+^{3-\sigma}q^2}\;\tn{always.}
\end{array}
\end{align}
Writing the denominator in the expression of $\c A_2$ as $\frac{4i\lambda k}{(\lambda^2{+}\bs q\bs q_+)^2{+}4k^2\lambda^2} = \rec{\lambda^2{+}\bs q\bs q_+{-}2ik\lambda}-\rec{\lambda^2{+}\bs q\bs q_+{+}2ik\lambda}$ and introducing new notation we thus have
\begin{align}
\label{e:A1def2}
&\quad\c A_1 = \frac k\pi\lim_{\lambda\to0}\slint{-1pt}{1pt}{}\m d^3\bs q\frac{f_1(\bs q)}{\lambda^2{+}q^2}\Bigg\{
\frac{\c G_{1+}}{\lambda^2{+}\bs q\bs q_+{-}2ik\lambda}-\nonumber\\
&\quad\qquad\qquad\qquad\qquad\qquad -\frac{\c G_{1-}}{\lambda^2{+}\bs q\bs q_-{+}2ik\lambda}\Bigg\},\\
&\quad\c A_2 = \frac k\pi\lim_{\lambda\to0}\slint{-1pt}{1pt}{}\m d^3\bs q\frac{f_2(\bs q)}{\lambda^2{+}q_+^2}\Bigg\{
\frac{\c G_{2a}\,{-}\,\c G_{2b}}{\lambda^2{+}\bs q\bs q_+{-}2ik\lambda}-\nonumber\\
&\quad\qquad\qquad\qquad\qquad\qquad -\frac{\c G_{2a}\,{+}\,\c G_{2b}}{\lambda^2{+}\bs q\bs q_+{+}2ik\lambda}\bigg\},
\label{e:A2def2}
\end{align}
where
\begin{align}
\label{e:Gdef1}
&\qquad \c G_{1-}=\c U_{1+}\c F_-(x_1),\\
&\qquad \c G_{1+}=\c U_{1+}\c F_-(x_1){-}2\c U_{1-}\c F_+(x_1),\\
&\qquad \c G_{2a}=\c U_{2+}\c F_-(x_2){-}\c U_{2-}\c F_+(x_2),\\
&\qquad \c G_{2b}=\c U_{2-}\c F_+(x_2).
\label{e:Gdef4}
\end{align}
We also introduce the following quantities, with corresponding indices in the notation: 
\begin{align}
\label{e:Hdef1}
&\qquad \c H_{1-} := G^*\c U_{1+},\\
&\qquad \c H_{1+} := {-}G\c U_{1-},\\
&\qquad \c H_{2a} := \rect2\big(G^*\c U_{2+}{-}\,G\c U_{2-}\big),\\
&\qquad \c H_{2b} := \rect2\big(G^*\c U_{2+}{+}\,G\c U_{2-}\big),
\label{e:Hdef4}
\end{align}
where $G\,{=}\,2\frac{\Gamma({-}2i\eta)}{\Gamma({-}i\eta)\Gamma(1{-}i\eta)}$ is the constant as defined in \Eq{e:Gdef}. These $\c H$s are also bounded; for simplicity, we can choose the $\c C$ constant already utilized to be such that for each $\c G$ and $\c H$ 
\begin{align}
\qquad |\c H|{\le}\c C,\qquad|\c G|{\le}\c C.
\end{align}
The usefulness of these $\c H$ quantities is that they ,,approximate'' the corresponding $\c G$ quantities around the singular point $x_{1,2}\,{=}\,1$. Concretely, it follows from various identities relating to the hypergeometric function that there are $\c C_1{>}0$ and $\c C'_1{>}0$ constants with which
\begin{align}
\label{e:Faround1}
&\qquad \big|\c U_+\big[\c F_-(x)\,{-}\,G^*\big]\big| \le\min\big\{\c C_1|1{-}x|,\c C_1'\big\},\\
&\qquad \big|\c U_-\big[\c F_+(x)\,{-}\,\rect2G\,{-}\,\frac{G^*/2}{(1{-}x)^{2i\eta}}\big]\big|\le\nonumber\\
&\qquad\qquad\qquad\qquad\qquad\le\min\big\{\c C_1|1{-}x|,\c C_1'\big\},
\end{align}
from which it follows (knowing that $\frac{\c U_{1,2-}}{\c U_{1,2+}}\,{=}\,(1{-}x_{1,2})^{2i\eta}$ and again the functional form of power functions) that for any $0{<}\gamma{<}1$ exponent there exists a $\c C_\gamma$ constant with which the following estimations (to be utilized below) hold true:
\begin{align}
\label{e:GHestim1}
&\qquad \big|\c G_{1\pm}{-}\c H_{1\pm}\big| \le \c C_\gamma|1{-}x_1|^\gamma,\\
&\qquad \big|\c G_{2a}{-}\c H_{2a}\big|     \le \c C_\gamma|1{-}x_2|^\gamma,\\
&\qquad \big|\c G_{2b}{-}\c H_{2b}\big|     \le \c C_\gamma|1{-}x_2|^\gamma.
\label{e:GHestim3}
\end{align}
Returning to \Eqs{e:A1def}{e:A2def}, we can leave the $\lambda^2$ terms from all the denominators (again; the difference of the ,,old'' and ,,new'' integrands is dominated by a $\lambda$-independent integrable function). We use two steps: first leave the $\lambda^2$s from next to $\bs q\bs q_\pm$, because (omitting the bounded $f_{1,2}\c G$ quantities)
\begin{align}
&\bigg|\rec{\bs q\bs q_\pm{\mp}2ik\lambda}{-}\rec{\lambda^2{+}\bs q\bs q_\pm{\mp}2ik\lambda}\bigg|\rec{\lambda^2{+}q^2}
=\frac{\lambda}{|\bs q\bs q_\pm{\mp}2ik\lambda|}\times\nonumber\\ &\quad\times\frac{\lambda}{|\lambda^2{+}\bs q\bs q_\pm{\mp}2ik\lambda|}\rec{\lambda^2{+}q^2}
\le\hspace{-1pt}\Bigg\{\hspace{-3pt}\begin{array}{l}
\rec{q^4}\;\;\tn{if $q{>}Q$,}\vspace{1pt}\\
\rec{4k^2q^2}\;\tn{always,}
\end{array}\\
&\bigg|\rec{\bs q\bs q_+{\pm}2ik\lambda}{-}\rec{\lambda^2{+}\bs q\bs q_+{\pm}2ik\lambda}\bigg|\rec{\lambda^2{+}q_+^2} =
\frac{\lambda}{|\bs q\bs q_+{\pm}2ik\lambda|}\times\nonumber\\ &\quad\times\frac{\lambda}{|\lambda^2{+}\bs q\bs q_+{\pm}2ik\lambda|}\rec{\lambda^2{+}q_+^2}
\,{\le}\Bigg\{\hspace{-3pt}\begin{array}{l}
\rec{q^4}\;\;\tn{if $q{>}Q$,}\vspace{1pt}\\
\rec{4k^2q_+^2}\;\tn{always,}
\end{array}
\end{align}
then leave the $\lambda^2$s next to $q^2$ and $q_+^2$, because
\begin{align*}
&\bigg|\frac{f_1(\bs q)}{\bs q\bs q_\pm{\mp}2ik\lambda}\bigg[\rec{q^2}{-}\rec{\lambda^2{+}q^2}\bigg]\bigg|\le
\frac{\lambda\,Kq^{\sigma-2}}{|\bs q\bs q_\pm{\mp}2ik\lambda|}\frac\lambda{\lambda^2{+}q^2}\le\nonumber\\&\qquad\le\Bigg\{\hspace{-3pt}\begin{array}{l}
\frac{K\lambda_0}{q^2\,q^{4-\sigma}}\;\;\tn{if $q{>}Q$,}\vspace{2pt}\\
\rec{2k}\frac{K}{2q^{3-\sigma}}\quad\tn{always,}
\end{array}\\
&\bigg|\frac{f_2(\bs q)}{\bs q\bs q_+{\pm}2ik\lambda}\bigg[\rec{q_+^2}{-}\rec{\lambda^2{+}q_+^2}\bigg]\bigg|\le
\frac{\lambda\,Kq_+^{\sigma-2}}{|\bs q\bs q_+{\pm}2ik\lambda|}\frac\lambda{\lambda^2{+}q_+^2}\le\nonumber\\&\qquad\le\Bigg\{\hspace{-3pt}\begin{array}{l}
\frac{K\lambda_0}{q^2\,q^{4-\sigma}}\;\;\tn{if $q{>}Q$,}\vspace{2pt}\\
\rec{2k}\frac{K}{2q_+^{3-\sigma}}\quad\tn{always.}
\end{array}
\end{align*}
We thus have so far, returning to real denominators,
\begin{align}
\label{e:A1def3}
&\quad \c A_1 = \frac k\pi\lim_{\lambda\to0}\slint{-1pt}{1pt}{}\m d^3\bs q\frac{f_1(\bs q)}{q^2}\bigg\{
\frac{2ik\lambda\c G_{1+}+\bs q\bs q_+\c G_{1+}}{(\bs q\bs q_+)^2{+}4k^2\lambda^2}+\nonumber\\
&\qquad\qquad\qquad\qquad\quad +\frac{2ik\lambda\c G_{1-}-\bs q\bs q_-\c G_{1-}}{(\bs q\bs q_-)^2{+}4k^2\lambda^2}\bigg\},\\
&\quad \c A_2 = \frac k\pi\lim_{\lambda\to0}\slint{-1pt}{1pt}{}\m d^3\bs q\frac{f_2(\bs q)}{q_+^2}
\frac{4ik\lambda\c G_{2a}-2\bs q\bs q_+\c G_{2b}}{(\bs q\bs q_+)^2{+}4k^2\lambda^2}.
\label{e:A2def3}
\end{align}
The terms in these integrands with $\lambda$s in the numerator converge pointwise to 0 everywhere except on the spherical $\bs q\bs q_\pm\,{=}\,0$ surfaces; however, the integral of these terms does not converge to 0. In this manner they are similar to the case when one approaches a Dirac delta with ever narrower and stronger peaks. The remaining terms in the integrands converge pointwise to forms that display $\bs q\bs q_\pm$ in the denominator, which is not integrable (around said surfaces; here they behave like $\sim\rec q_\Delta$ where $q_\Delta$ is the distance of the $\bs q$-point from the spheres): their integral-limits are to be expressed similarly to that of a (Cauchy) principal value integral. These considerations, together with calculating the level surfaces of $x_{1,2}$ in the $\lambda\,{\to}\,0$ case, suggest the introduction of the new coordinate systems: for $\c A_1$, the $a\,{\in}\,\B R$, $\beta\,{\in}\,\B R_0^+$, $\varphi\,{\in}\,[{-}\pi,\pi]$ coordinates, with which the expression of $\bs q$, denoted here below by $\bs q^{(1)}(a,\beta,\varphi)$, is specified in \Eq{e:A1coord} and illustrated on Fig.~\ref{f:A1coord}, and for $\c A_2$, the $b\,{\in}\,\B R_0^+$, $y\,{\in}\,[{-}1,1]$, $\varphi\,{\in}\,[{-}\pi,\pi]$ coordinates, with which the expression of $\bs q$, denoted by $\bs q^{(2)}(b,y,\varphi)$, is specified in \Eq{e:A2coord} and illustrated on Fig.~\ref{f:A2coord}. We do not repeat the coordinate definitions, \Eqs{e:A1coord}{e:A2coord} here, just state that for $\lambda\,{\to}\,0$, 
\begin{align}
\quad x_1\;\stackrel{\lambda\to0}\longrightarrow\;\rec{a^2}-\m{sgn}(a)\cdot i0,\quad\; x_2\;\stackrel{\lambda\to0}\longrightarrow\;\frac{4b}{(1{+}b)^2},
\end{align}
where $i0$ denotes which side the $[1,\infty]$ branch cut of the $\c F_\pm$ function is to be approached from. For simplicity, for a while from now on we suppress the $\varphi$ variable (as if the $f_{1,2}$ functions were axially symmetric in $\varphi$): everything is to be understood as having also $\varphi$ as a variable, and the integral over $\varphi$ is written up as a $2\pi$ factor. Instead of $\lambda$, we use the dimensionless $\Lambda$ as
\begin{align}
\quad \lambda = 2k\Lambda,\quad\lambda_0 = 2k\Lambda_0;
\end{align}
the allowed range is $\Lambda\,{<}\,\Lambda_0\,{<}\,1$, owing to $\lambda\,{<}\,\lambda_0\,{<}\,2k$. We thus write up our expressions of $\c A_1$ and $\c A_2$ in the new coordinates, paying attention to the Jacobian, as
\begin{align}
\label{e:A1def4}
&\quad\c A_1\,{=}\,\lim_{\Lambda\to0}\sintl{-2pt}{-5pt}{-\infty}\infty\m da\sintl{-2pt}{-3pt}0\infty\m d\beta\,\Bigg\{
\frac{i\Lambda+\frac{a(a{+}1)\beta^2}{a^2{+}\beta^2}}{\Lambda^2{+}\frac{a^2(a{+}1)^2\beta^4}{(a^2{+}\beta^2)^2}}\c G_{1+} +\nonumber\\
&\qquad\qquad + \frac{i\Lambda-\frac{a(a{-}1)\beta^2}{a^2{+}\beta^2}}{\Lambda^2{+}\frac{a^2(a{-}1)^2\beta^4}{(a^2{+}\beta^2)^2}}\c G_{1-}\Bigg\}
\frac{a^2\beta\,f_1(a,\beta)}{(a^2{+}\beta^2)^2},\\
&\quad\c A_2\,{=}\,\lim_{\Lambda\to0}\sintl{-1pt}{-3pt}{-1}1\m dy\sintl{-2pt}{-4pt}0\infty\m db\,\Bigg\{
\frac{i\Lambda\c G_{2a}}{\Lambda^2{+}\frac{(b{-}1)^2(1{-}y^2)^2}{16(1{+}by^2)^2}}-\nonumber\\
&\qquad\qquad-\frac{\frac{(b{-}1)(1{-}y^2)}{4(1{+}by^2)}\c G_{2b}}{\Lambda^2{+}\frac{(b{-}1)^2(1{-}y^2)^2}{16(1{+}by^2)^2}}\Bigg\}\frac{(1{-}y)^2\,f_2(b,y)}{2(1{+}by^2)^2},
\label{e:A2def4}
\end{align}
with the $\c G$s understood here as having been expressed as functions of the new coordinates, and with the abbreviated notations $f_1(a,\beta)$ and $f_2(b,y)$ used instead of $f_1(\bs q^{(1)}(a,\beta,\varphi))$ and $f_2(\bs q^{(2)}(b,y,\varphi))$, respectively.

The ,,critical'' surfaces are the $a\,{=}\,\pm1$ and the $b\,{=}\,1$ spheres: they are the support of the terms with $\Lambda$s in the numerator. In these terms we thus try to substitute $f_1(a,\beta)$ with $f_1(\pm1,\beta)$ and $f_2(b,y)$ with $f_2(1,y)$. To justify this, we need the easily verifiable estimations
\begin{align}
\label{e:estim:T1shellQQ}
& \quad \big|\bs q(a,\beta){-}\bs q({\pm}1,\beta)\big| \le \frac{2\sqrt2k\beta^2|a{\mp}1|}{(1{+}\beta)\sqrt{a^2{+}\beta^2}},\\
& \quad \big|\bs q(b,y){-}\bs q(1,y)\big| \le \frac{k(1{-}y^2)|\sqrt b{-}1|}{\sqrt{1{+}by^2}},\\
\label{e:estim:T2shellQQ}
& \quad \big|\bs q({\pm}1,\beta)\big| \le \frac{2\sqrt2k\beta}{1{+}\beta},\\
& \quad \big|\bs q_+(1,y)\big| \le \sqrt2k(1{+}y),
\end{align}
which, combined with Eqs.~\r{e:estim:f2K1}--\r{e:estim:f2K2}, mean that 
\begin{align}
&\quad \big|f_1(a,\beta){-}f_1(\pm1,\beta)\big|\le\nonumber\\
&\qquad\qquad \le \m{min}\bigg\{\frac{K_1\beta^{2\sigma}|a{\mp}1|^\sigma}{(1{+}\beta)^\sigma(a^2{+}\beta^2)^{\sigma/2}},K'\bigg\},\\
&\quad \big|f_2(b,y){-}f_2(1,y)\big|, \le\nonumber\\
&\qquad\qquad \le\m{min}\bigg\{\frac{K_2(1{-}y^2)^\sigma|\sqrt b{-}1|^\sigma}{(1{+}by^2)^{\sigma/2}},K'\bigg\},\\
&\quad \big|f_1(\pm1,\beta)\big|\le K_1\frac{\beta^\sigma}{(1{+}\beta)^\sigma}, \label{e:f1qq}\\
&\quad\big|f_2(1,y)\big|\le \overline K_2(1{+}y)^\sigma,
\label{e:f2qq}
\end{align}
where $K_1\,{=}\,k^\alpha2^{3\alpha/2}K$, $K_2\,{=}\,k^\alpha K$, and $\overline K_2\,{=}\,2^{\alpha/2}K_2$. 

With these, justification of the $f_1(a,\beta)\,{\to}\,f_1({\pm}1,\beta)$ and the $f_2(b,y)\,{\to}\,f_2(1,y)$ substitutions in the terms of \Eqs{e:A1def4}{e:A2def4} with $\Lambda$s in the numerators again relies on that the difference of the two integrands is overestimated by a $\Lambda$-independent integrable function (integrability is best established by separately investigating various blocks of the two-dimensional integration domain): putting together our various estimations and \Eq{e:lorentzestim}, we have
\begin{align}
\label{e:estim1A}
&\quad \bigg|\frac{\c G_\pm a^2\beta}{(a^2{+}\beta^2)^2}\frac{i\Lambda[f_1(a,\beta){-}f_1(\mp1,\beta)]}{\Lambda^2{+}\frac{a^2(a{\pm}1)^2\beta^4}{(a^2{+}\beta^2)^2}}\bigg|
\le \rec{2\beta}\frac{\c C|a|}{a^2{+}\beta^2}\times\nonumber\\
&\qquad\quad\times\m{min}\bigg\{\frac{K_1\beta^{2\sigma}|a{\mp}1|^{\sigma-1}}{(1{+}\beta)^\sigma(a^2{+}\beta^2)^{\sigma/2}},
\frac{K'}{|a{\mp}1|}\bigg\},\\
&\quad \bigg|\frac{(1{-}y)^2\,\c G_{2a}}{2(1{+}by^2)^2}\frac{i\Lambda[f_2(b,y){-}f_2(1,y)]}{\Lambda^2{+}\frac{(b{-}1)^2(1{-}y^2)^2}{16(1{+}by^2)^2}}\bigg|
\le\frac{\c C}{|b{-}1|}\frac{1{-}y}{1{+}by^2}\times\nonumber\\
&\qquad\quad\times\m{min}\bigg\{\frac{K_2(1{-}y)^\sigma|\sqrt b{-}1|^\sigma}{(1{+}by^2)^{\sigma/2}(1{+}y)^{1-\sigma}},\frac{K'}{(1{+}y)}\bigg\}.
\label{e:estim2A}
\end{align}
In these same terms in \r{e:A1def4} and \r{e:A2def4}, the ones with $\Lambda$s in the numerator, we can also substitute the $\c G$s with the corresponding $\c H$s (that ,,approximate'' them around the critical $x_{1,2}\,{=}\,1$, i.e. $a\,{=}\,{\pm}1$ and $b\,{=}\,1$ spheres; see Eqs.~\r{e:Gdef1}--\r{e:Gdef4} and \r{e:Hdef1}--\r{e:Hdef4} for the definitions). For such difference of the integrands, the fact that $1{-}x_{1,2}$ tends to 0 around the critical spheres together with Eqs.~\r{e:GHestim1}--\r{e:GHestim3} help to construct an integrable dominant function. Concretely, if $\gamma{<}1$, then
\begin{align}
&\quad\big|1{-}x_1\big|^\gamma = \Bigg|\frac{\frac{a(a{\pm}1)\beta^2}{a^2{+}\beta^2}{+}\Lambda^2{\mp}i\Lambda}{\frac{a^2\beta^2}{a^2{+}\beta^2}{+}\Lambda^2}
\cdot\frac{\frac{a(a{\mp}1)\beta^2}{a^2{+}\beta^2}{+}\Lambda^2{\pm}i\Lambda}{\frac{a^2\beta^2}{a^2{+}\beta^2}{+}\Lambda^2}\Bigg|^\gamma\hspace{-3pt}\le\nonumber\\
&\quad\le\bigg[\frac{|a{\pm}1|}{|a|}{+}\frac{2\Lambda(a^2{+}\beta^2)}{a^2\beta^2}\bigg]^\gamma{\cdot}\,
\bigg[2\frac{(|a|{+}1)(\beta{+}1)}{|a|\beta}\bigg]^\gamma\le\nonumber\\
&\quad\le 2\frac{(|a|{+}1)^\gamma(\beta{+}1)^\gamma}{|a|^{2\gamma}\beta^\gamma}
\bigg[|a{\pm}1|^\gamma{+}\Lambda^\gamma\frac{(a^2{+}\beta^2)^\gamma}{|a|^\gamma\beta^{2\gamma}}\bigg],
\label{e:1mx1estim}\\
&\quad|1{-}x_2|^{\gamma} \,{=}\,\Bigg|\frac{8\Lambda^2\big(2\Lambda^2{+}\frac{(b{+}1)(1{-}y^2)}{1{+}by^2}\big){+}\frac{(b{-}1)^2(1{-}y^2)^2}{(1{+}by^2)^2}}
{\big[4\Lambda^2{+}\frac{(1{+}b)(1{-}y)^2}{1+by^2}\big]\big[4\Lambda^2{+}\frac{(1{+}b)(1{+}y)^2}{1+by^2}\big]}\Bigg|^{\gamma}\le\nonumber\\
&\quad\le\bigg|\frac{9\Lambda^2}{b{+}1}\frac{1{+}by^2}{1{-}y^2}{+}\frac{(b{-}1)^2}{(b{+}1)^2}\bigg|^{\gamma}
\le\bigg|\frac{9\Lambda}{b{+}1}\frac{1{+}by^2}{1{-}y^2}{+}\frac{b{-}1}{b{+}1}\bigg|^{\gamma}\le\nonumber\\
&\quad\le \frac{9^\gamma\Lambda^\gamma(1{+}by^2)^\gamma}{(b{+}1)^\gamma(1{-}y^2)^{\gamma}}{+}\frac{|b{-}1|^\gamma}{(b{+}1)^\gamma};
\label{e:1mx2estim}
\end{align}
with these and Eqs.\@ \r{e:GHestim1}, \r{e:GHestim3}, the estimations on $f_{1,2}$, and adroit usage of \Eq{e:lorentzestim2} we arrive at
\begin{align}
&\quad\bigg|\frac{f_1({\mp}1,\beta)}{(a^2{+}\beta^2)^2}\frac{i\Lambda[\c G_{1\pm}{-}\c H_{1\pm}]a^2\beta}{\Lambda^2{+}\frac{a^2(a{\pm}1)^2\beta^4}{(a^2{+}\beta^2)^2}}\bigg|
\le\nonumber\\
&\quad\le\frac{|a|^{2-2\gamma}(|a|{+}1)^\gamma}{(a^2{+}\beta^2)^2}\frac{\beta^{1+\sigma-\gamma}}{(\beta{+}1)^{\sigma-\gamma}}
\frac{2K_1\c C_1\Lambda|a{\pm}1|^\gamma}{\Lambda^2{+}\frac{a^2(a{\pm}1)^2\beta^4}{(a^2{+}\beta^2)^2}}+\nonumber\\
&\qquad+\frac{(|a|{+}1)^\gamma|a|^{2-3\gamma}}{(a^2{+}\beta^2)^{2-\gamma}}
\frac{\beta^{1+\sigma-3\gamma}}{(\beta{+}1)^{\sigma-\gamma}}\frac{2K_1\Lambda^{1+\gamma}\c C_1}{\Lambda^2{+}\frac{a^2(a{\pm}1)^2\beta^4}{(a^2{+}\beta^2)^2}}\le\nonumber\\
&\quad\le\frac{K_1\c C_\gamma(|a|{+}1)^\gamma|a|^{1-2\gamma}}{(a^2{+}\beta^2)^2(\beta{+}1)^{\sigma-\gamma}}\frac{|a{\pm}1|^{\gamma-1}}{\beta^{1+\gamma-\sigma}}
\bigg[1{+}\rec{2^\gamma}\bigg],\label{e:estim1B}\\
&\quad \bigg|\frac{f_2(1,y)}{2(1{+}by^2)^2}\frac{i\Lambda(1{-}y)^2[\c G_{2a}{-}\c H_{2a}]}{\Lambda^2{+}\frac{(b{-}1)^2(1{-}y^2)^2}{16(1{+}by^2)^2}}\bigg| \le\nonumber\\
&\quad\le\frac{\overline K_2(1{+}y)^{\sigma-\gamma}(1{-}y)^{2-\gamma}}{2(1{+}by^2)^{2-\gamma}(b{+}1)^\gamma}
\frac{9^\gamma\Lambda^{1+\gamma}\c C_\gamma}{\Lambda^2{+}\frac{(b{-}1)^2(1{-}y^2)^2}{16(1{+}by^2)^2}}+\nonumber\\
&\qquad+\frac{\overline K_2(1{+}y)^\sigma}{2(1{+}by^2)^2}\frac{\Lambda(1{-}y)^2\c C_\gamma}{\Lambda^2{+}\frac{(b{-}1)^2(1{-}y^2)^2}{16(1{+}by^2)^2}}
\frac{|b{-}1|^\gamma}{(b{+}1)^\gamma}\le\nonumber\\
&\quad\le\frac{\overline K_2\c C_\gamma(1{+}y)^{\sigma-1}(1{-}y)}{(1{+}by^2)(b{+}1)^\gamma|b{-}1|^{1-\gamma}}\bigg[\frac{9^\gamma}{8^\gamma}{+}1\bigg].
\label{e:estim2B}
\end{align}
The obtained dominants are integrable (verified by separately paying attention to the behavior at infty and around $a\,{=}\,{\pm}1$ of $b{=}1$) if $0{<}\gamma{<}\sigma$; we can choose such $\gamma$, so the investigated interchange of the integrands is indeed allowed. At this point we have
\begin{align}
\label{e:A1def6}
&\quad\c A_1 = \lim_{\Lambda\to0}\sint{-2pt}{-10pt}0\infty\m d\beta\sint{-2pt}{-12pt}{-\infty}\infty\m da\Bigg\{
\frac{f_1(a,\beta)}{(a^2{+}\beta^2)^3}\frac{\c G_{1+}a^3\beta^3(a{+}1)}{\Lambda^2{+}\frac{a^2(a{+}1)^2\beta^4}{(a^2{+}\beta^2)^2}}
+\nonumber\\&\quad+\frac{i\Lambda a^2\beta}{(a^2{+}\beta^2)^2}\Bigg[\frac{\c H_{1+}f_1(-1,\beta)}{\Lambda^2{+}\frac{a^2(a{+}1)^2\beta^4}{(a^2{+}\beta^2)^2}}
+\frac{\c H_{1-}f_1(1,\beta)}{\Lambda^2{+}\frac{a^2(a{-}1)^2\beta^4}{(a^2{+}\beta^2)^2}}\Bigg]
-\nonumber\\&\quad -\frac{f_1(a,\beta)}{(a^2{+}\beta^2)^3}\frac{\c G_{1-}a^3\beta^3(a{-}1)}{\Lambda^2{+}\frac{a^2(a{-}1)^2\beta^4}{(a^2{+}\beta^2)^2}}\Bigg\},\\
&\quad\c A_2 = \lim_{\Lambda\to0}\sint{-2pt}{-9pt}{-1}1\m dy\sint{-2pt}{-12pt}0\infty\m db\Bigg\{\frac{f_2(1,y)}{2(1{+}by^2)^2}
\frac{i\Lambda\c H_{2a}(1{-}y)^2}{\Lambda^2{+}\frac{(b{-}1)^2(1{-}y^2)^2}{16(1{+}by^2)^2}}-\nonumber\\&\quad- \frac{f_2(b,y)}{8(1{+}by^2)^3}
\frac{\c G_{2b}(1{-}y)^2(1{-}y^2)(b{-}1)}{\Lambda^2{+}\frac{(b{-}1)^2(1{-}y^2)^2}{16(1{+}by^2)^2}}\Bigg\}.
\label{e:A2def6}
\end{align}
After some deliberation, it turns out that in the terms with $\Lambda$s in the numerators the following substitutions can and should be performed (still because these terms have the $a\,{=}\,{\pm}1$ and $b\,{=}\,1$ spheres as supports):
\begin{align}
\label{e:A1subst}
&\quad\frac{f_1(\mp 1,\beta)}{a^2{+}\beta^2}\;\to\;\frac{\mp2i\,f(\mp 1,\beta)}{(a{+}i)^2(1{+}\beta^2)},\quad\tn{and}\\
&\quad\frac{f_2(1,\beta)}{2(1{+}by^2)}\;\to\;\frac{f_2(1,\beta)}{(1{+}y^2)(b{+}1)};
\label{e:A2subst}
\end{align}
the justifications being again the following estimations which provide a $\Lambda$-independent integrable dominant for the difference of the original and the desired integrands:
\begin{align}
&\quad\bigg|\frac{a^2\beta}{(a^2{+}\beta^2)^2}\frac{i\Lambda\c H_{1\pm}\,f_1({\mp}1,\beta)}{\Lambda^2{+}\frac{a^2(a{\pm}1)^2\beta^4}{(a^2{+}\beta^2)^2}}
\bigg[1{-}\frac{\mp2i}{(a{+}i)^2}\frac{a^2{+}\beta^2}{1{+}\beta^2}\bigg]\bigg| \le\nonumber\\&\qquad{\le}
\frac{K_1\c C|a|\beta^{\sigma-1}\big|(\beta^2{-}1)(a{\pm}1){\pm}2i(a{\pm}\beta^2)\big|}{2(\beta{+}1)^\sigma(\beta^2{+}1)(a{+}i)^2(a^2{+}\beta^2)},\label{e:estim1C}\\
&\quad\bigg|\frac{(1{-}y)^2}{2(1{+}by^2)^2}\frac{i\Lambda\c H_{2a}\,f_2(1,y)}{\Lambda^2{+}\frac{(b{-}1)^2(1{-}y^2)^2}{16(1{+}by^2)^2}}
\bigg[1{-}\frac{2(1{+}by^2)}{(1{+}y^2)(1{+}b)}\bigg]\bigg| \le \nonumber\\&\qquad{\le}
\frac{(1{-}y)^2\c CK'}{(1{+}by^2)(1{+}y^2)(b{+}1)}.
\label{e:estim2C}
\end{align}
The next step is to subtract terms similar to the lastly obtained ones (with $\Lambda$s in the numerators) from the others, and add them back in a way that they can be combined to have a simple ,,pole structure''; after some experimentation, one can obtain the following expressions:
\begin{align}
\quad\c A_1 = \c A_{1\c P}+\c A_{1\delta},\qquad
\c A_2 = \c A_{2\c P}+\c A_{2\delta},
\end{align}
where in the $\delta$-terms one can put back the original \r{e:Hdef1}--\r{e:Hdef4} definitions of the $\c H$ quantities, and also use the relation $\c U_{2+}^*\,{=}\,\c U_{2-}$, to arrive at
\begin{align}
\label{e:A1def7cP}
&\c A_{1\c P} \,{=}\lim_{\Lambda\to0}\sintl{-3pt}{-6pt}{-\infty}\infty\m da\sintl{-3pt}{-3pt}0\infty\m d\beta\biggg\{\hspace{-2pt}
(a{+}1)\frac{\frac{\c G_{1+}f_1(a,\beta)}{a^2{+}\beta^2}{+}\frac{2i\c H_{1+}f_1({-}1,\beta)}{(a{+}i)^2(\beta^2{+}1)}}
{\Lambda^2{+}\frac{a^2(a{+}1)^2\beta^4}{(a^2{+}\beta^2)^2}}-\nonumber\\
&\quad {-}(a{-}1)\frac{\frac{\c G_{1-}f_1(a,\beta)}{a^2{+}\beta^2}{-}\frac{2i\c H_{1-}f_1(1,\beta)}{(a{+}i)^2(\beta^2{+}1)}}
{\Lambda^2{+}\frac{a^2(a{-}1)^2\beta^4}{(a^2{+}\beta^2)^2}}\hspace{-2pt}\biggg\}\frac{a^3\beta^3}{(a^2{+}\beta^2)^2},\\
\label{e:A1def7delta}
&\c A_{1\delta}\,{=}\,\lim_{\Lambda\to0}\sint{-2pt}{-12pt}{-\infty}\infty\m da\sint{-2pt}{-10pt}0\infty\m d\beta\bigg\{
\frac{G\,\c U_{1-}f_1({-}1,\beta)}{a(a{+}1)\beta^2{-}i\Lambda(a^2{+}\beta^2)} - \nonumber\\
&\quad-\frac{G^*\c U_{1+}f_1(1,\beta)}{a(a{-}1)\beta^2{+}i\Lambda(a^2{+}\beta^2)}\bigg\}\frac{2ia^2\beta}{(a{+}i)^2(\beta^2{+}1)},\\
\label{e:A2def7cP}
&\c A_{2\c P} = \lim_{\Lambda\to0}\sintl{-2pt}{-4pt}{-1}1\m dy\sintl{-3pt}{-4pt}0\infty\m db\frac{(1{-}y)^2}{8(1{+}by^2)^2}
\frac{(b{-}1)(1{-}y)^2}{\Lambda^2{+}\frac{(b{-}1)^2(1{-}y^2)^2}{16(1{+}by^2)^2}}\times\nonumber\\
&\qquad\qquad\qquad\times\bigg[\frac{2\c H_{2b}f_2(1,y)}{(1{+}b)(1{+}y^2)}{-}\frac{\c G_{2b}f_2(b,y)}{1{+}by^2}\bigg],\\
&\c A_{2\delta} = {-}\hspace{-2pt}\lim_{\Lambda\to0}\sint{-2pt}{-9pt}{-1}1\m dy\sint{-2pt}{-12pt}0\infty\m db
\,\m{Re}\frac{\frac{G^*\c U_{2+}}{1{+}by^2}\frac{(1{-}y)^2f_2(1,y)}{(b{+}1)(1{+}y^2)}}{\frac{(b{-}1)(1{-}y^2)}{4(1{+}by^2)}{+}i\Lambda}.
\label{e:A2def7delta}
\end{align}
This formulation pre-supposes the statement that the limits of the $\c P$-terms (resembling to principal value integrals, centered on the $a\,{=}\,{\pm}1$ and $b\,{=}\,1$ spheres) and the $\delta$-terms (resembling to Dirac deltas as their support, these spheres, are lower dimensional) exist separately.

In the $\c P$-terms, we can finally interchange the limit and the integral; again by virtue of Lebesgue's theorem. At this point we defer writing up explicity the required dominant functions; they can be best constructed in three steps, roughly corresponding to the steps in the simplifications of the $\Lambda$-containing terms above. Owing to the triangle inequality, we have for the $\c A_{\c P}$ term
\begin{align}
&\;\;\bigg|\frac{\c G_{1\pm}f_1(a,\beta)}{a^2{+}\beta^2}{\pm}\frac{2i\,\c H_{1\pm}f_1(\mp1,\beta)}{(a{+}i)^2(\beta^2{+}1)}\bigg|\le\nonumber\\
&\;\;\;{\le}\,\bigg|\c G_{1\pm}\frac{f_1(a,\beta){-}f_1(\mp1,\beta)}{a^2{+}\beta^2}\bigg|
\,{+}\,\bigg|f_1(\mp1,\beta)\frac{\c G_{1\pm}{-}\c H_{1\pm}}{a^2{+}\beta^2}\bigg|+\nonumber\\
&\quad+\bigg|f_1(\mp1,\beta)\c H_{1\pm}\bigg[\rec{a^2{+}\beta^2}{\pm}\frac{2i}{(a{+}i)^2(\beta^2{+}1)}\bigg]\bigg|
\end{align}
while for the $\c A_{2\c P}$ term, 
\begin{align}
&\;\;\bigg|\frac{\c G_{2b}f_2(b,y)}{1{+}by^2}{-}\frac{2\,\c H_{2b}f_2(1,y)}{(1{+}b)(1{+}y^2)}\bigg|\le\nonumber\\
&\;\;\;{\le}\,\bigg|\c G_{2b}\frac{f_2(b,y){-}f_2(1,y)}{1{+}by^2}\bigg|\,{+}\,\bigg|f_2(1,y)\frac{\c G_{2b}{-}\c H_{2b}}{1{+}by^2}\bigg|+\nonumber\\
&\quad+\bigg|f_2(1,y)\c H_{2b}\bigg[\rec{1{+}by^2}{-}\frac{2}{(1{+}b)(1{+}y^2)}\bigg]\bigg|,
\end{align}
and one can write up integrable dominant functions for the three terms separately, with estimations very similar to those used around Eqs.~\r{e:estim1A}--\r{e:estim2A}, \r{e:estim1B}--\r{e:estim2B} and \r{e:estim1C}--\r{e:estim2C}, sometimes instead of using \Eq{e:lorentzestim}, just omitting $\Lambda^2$ in the denominator. (The purpose of the manipulations so far were precisely to arrive at integrals where such exchange of limits and integrals is possible.) Performing and simplifying the limit in this way, putting back the omitted $\varphi$ integration variable, and (making use of $f$ and hence $f_1$ being an even function) substituting $a\,{\to}\,{-}a$ in one half of $\c A_{1\c P}$, and also using \Eqs{e:Fidentity1}{e:Fidentity2}, one arrives at the expressions of $\c A_1$ and $\c A_{2\c P}$ in \Eqs{e:A1final}{e:A2Pfinal}.

In the $\c A_{1\delta}$ and $\c A_{2\delta}$ terms, we can explicitly calculate the integrals that do not involve the $f$ function, i.e. the ones over $a$ and $b$, respectively; the $\c A_{1\delta}$ term turns out to yield zero.%
\footnote{
As we will see soon here, this is essentially the consequence of the substitution made in \Eq{e:A1subst}. It certainly would have been possible to make a similarly advantageous substitution in $\c A_2$ (that on top of being analytically calculable, yields zero in the separate $\c A_{2\delta}$-like terms); after some attempts, we gave up the search for such. The ,,effectivity'' of the final numerical integral depends on this only very marginally.
}
Proceeding from the forms displayed in \Eqs{e:A1def7delta}{e:A2def7delta}, the first step is to substitute the $\c U_{1,2\pm}$ quantities with the following simpler $\c V_{1,2\pm}$ ones, with the $\Lambda^2$s omitted:
\begin{align}
&\quad\c V_{1\pm}:=\frac{(\lambda^2{+}q^2)^{\pm 2i\eta}}{\z{\bs q\bs q_\mp{\pm}2ik\lambda}^{\pm 2i\eta}},\\
&\quad\c V_{2\pm}:=\frac{((\lambda^2{+}q^2)(\lambda^2{+}q_+^2))^{\pm i\eta}}{\z{\bs q\bs q_+{\pm}2ik\lambda}^{\pm 2i\eta}},
\end{align}
which is justified by proving that (omitting constants)
\begin{align}
\label{e:UV1}
&\sint{-2pt}{-12pt}{-\infty}\infty\m da\sint{-2pt}{-12pt}0\infty\m d\beta
\frac{\frac{a^2\beta}{a^2{+}\beta^2}\big(\c U_{1\pm}{-}\c V_{1\pm}\big)f_1(\pm1,\beta)}
{(a{+}i)^2(\beta^2{+}1)\big[\frac{a(a{\mp}1)\beta^2}{a^2{+}\beta^2}{\pm}i\Lambda\big]} \stackrel{\Lambda\to0}\longrightarrow 0,\\
&\sint{-2pt}{-9pt}{-1}1\m dy\sint{-2pt}{-12pt}0\infty\m db
\frac{\frac{(1{-}y)^2}{1{+}by^2}\big(\c U_{2\pm}{-}\c V_{2\pm}\big)f_2(1,y)}{(1{+}b)(1{+}y^2)\big[\frac{(b{-}1)(1{-}y^2)}{4(1{+}by^2)}{\pm}i\Lambda\big]}
\stackrel{\Lambda\to0}\longrightarrow 0.
\label{e:UV2}
\end{align}
This is true because%
\footnote{
Here we use the fact that for $x\,{\in}\,\B R$ and $z\,{\in}\,\B C$, $|z^{ix}|\le e^{\pi|x|}$, and that if $|z|\,{\le}\,R{<}1$, then
$\big|(1{+}z)^{ix}\,{-}\,1\big|\,{\le}\,\rec{1{-}R}e^{\pi|x|}\cdot|x||z|$; for the sake of this latter one had we had to assume that $\Lambda\,{\le}\,\Lambda_0\,{<}\,1$.
}
\begin{align}
&\quad|\c U_{1\pm}{-}\c V_{1\pm}| = \Bigg|\frac{\big(1{+}\frac{\lambda^2}{\bs q\bs q_\mp{\pm}2ik\lambda}\big)^{\mp2i\eta}\,{-}\,1}
{(\bs q\bs q_\mp{\pm}2ik\lambda)^{\pm2i\eta}(\lambda^2{+}q^2)^{\mp2i\eta}}\Bigg| \le\nonumber\\&\quad \le
e^{2\pi\eta}\frac{2\eta e^{2\pi\eta}}{1{-}\Lambda_0}\Big|\frac{\lambda^2}{\bs q\bs q_\mp{\pm}2ik\lambda}\Big| \le 
\frac{2\eta e^{4\pi\eta}}{1{-}\Lambda_0}\Lambda,
\end{align}
and very similarly we also obtain $|\c U_{2\pm}{-}\c V_{2\pm}|\,{\le}\,L\Lambda$ for an appropriate $L$ constant; so the integrands in \r{e:UV1} and \r{e:UV2} converge to 0 pointwise almost everywhere, and they are easily dominated by integrable functions. Our expressions of $\c A_{1\delta}$ and $\c A_{2\delta}$ are now
\begin{align}
&\quad\c A_{1\delta}\,{=}\,2i\big(G\c A_{1\delta-}\,{-}\,G^*\c A_{1\delta+}\big),\\
&\quad\c A_{2\delta}\,{=}\,{-}2\big(G\c A_{2\delta-}\,{+}\,G^*\c A_{2\delta+}\big),
\label{e:A2dA2dpm}
\end{align}
where the auxiliary quantities, written up (together with $\c V_{1,2\pm}$) in the utilized coordinates are
\begin{align}
&\quad\c A_{1\delta\pm}= \lim_{\Lambda\to0}\sint{-2pt}{-11pt}0\infty\m d\beta\frac{f_1(\pm1,\beta)}{\beta^2{+}1}
\sint{-2pt}{-13pt}{-\infty}\infty\m da\Bigg\{\frac{a^2\beta}{(a{+}i)^2}{\times}\nonumber\\
&\qquad\qquad\qquad\qquad{\times}\frac{\big((a{-}B_u)(a{-}B_u^*)\big)^{\pm2i\eta}}{\big((a{-}B_0)(a{-}B_1)\big)^{1\pm2i\eta}}\Bigg\},
\end{align}
with the singularities being
\begin{align}
B_u{=}\frac{i\beta\Lambda}{\sqrt{\beta^2{+}\Lambda^2}},\quad
\begin{array}{l}
B_0=\frac{\beta^2{-}\sqrt{\beta^4(1{\mp}4i\Lambda){+}4\Lambda^2\beta^2}}{{\pm}2(\beta^2{\pm}i\Lambda)},\vspace{2pt}\\
B_1=\frac{\beta^2{+}\sqrt{\beta^4(1{\mp}4i\Lambda){+}4\Lambda^2\beta^2}}{{\pm}2(\beta^2{\pm}i\Lambda)},
\end{array}
\end{align}
as well as
\begin{align}
&\quad\c A_{2\delta\pm}=\lim_{\Lambda\to0}\sint{-2pt}{-9pt}{-1}1\m dy\,f_2(1,y)\sint{-2pt}{-12pt}0\infty\m db\Bigg\{\c K(y){\times}\nonumber\\
&\qquad\qquad\qquad\qquad{\times}\frac{(b{+}B_l)^{\pm i\eta}(b{+}B_r)^{\pm i\eta}}{(b{+}1)(b{-}B'_1)^{1\pm2i\eta}}\Bigg\},
\label{e:A3dpm}
\end{align}
with the singularities and the $\c K(y)$ function being
\begin{align}
&\quad B_l\,{=}\,\frac{1{+}\frac{4\Lambda^2}{(1{-}y)^2}}{1{+}\frac{4\Lambda^2y^2}{(1{-}y)^2}},\quad
B_r\,{=}\,\frac{1{+}\frac{4\Lambda^2}{(1{+}y)^2}}{1{+}\frac{4\Lambda^2y^2}{(1{+}y)^2}},\quad
B'_1\,{=}\,\frac{1{\mp}\frac{4i\Lambda}{1{-}y^2}}{1{\pm}\frac{4i\Lambda y^2}{1{-}y^2}},\nonumber\\
&\quad\c K(y)\,{=}\,\frac{1{-}y}{1{+}y}\frac{\big[1{+}\frac{4\Lambda^2y^2}{(1{-}y)^2}\big]^{\pm i\eta}\big[1{+}\frac{4\Lambda^2y^2}{(1{+}y)^2}\big]^{\pm i\eta}}
{(1{+}y^2)\big[1{\pm}\frac{4i\Lambda y^2}{1{-}y^2}\big]^{1\pm2i\eta}}.
\end{align}
In case of $\c A_{1\delta\pm}$, we can state the following:
\begin{align}
&\quad\begin{array}{l}
\m{Re}(B_0){\lessgtr}0,\\
\m{Im}(B_0){>}0,\\
\lim\limits_{\Lambda\to0}B_0=0,
\end{array}\qquad
\begin{array}{l}
\m{Re}(B_1){\gtrless}0,\\
\m{Im}(B_1){<}0,\\
\lim\limits_{\Lambda\to0}B_1=\pm1,
\end{array}
\end{align}
and based on this, investigation of the phases of the complex quantities leads us to the possibility to rewrite the integrals in the following form (with branch cuts going in the away direction from the real axis):  
\begin{align*}
&\quad \c A_{1\delta\pm}= \lim_{\Lambda\to0}\sint{-2pt}{-11pt}0\infty\m d\beta\frac{f_1(\pm1,\beta)}{\beta^2{+}1}\sint{-2pt}{-13pt}{-\infty}\infty\m da\Bigg\{
\frac{a^2\beta}{(a{+}i)^2}\times\nonumber\\&\qquad\times\frac{(ia{-}iB_u)^{\pm2i\eta}(iB_d{-}ia)^{\pm2i\eta}}{(ia{-}iB_0)^{1\pm2i\eta}(iB_1{-}ia)^{1\pm2i\eta}}\Bigg\}=(*),
\end{align*}
and then (also knowing the $\sim\rec{a^2}$ decrease at infinity) to tilt the integration path on the $a$ plane by an (acute) angle $\pm\phi$ so that this new (straight) path (denoted by $\gamma$) does not intersect any of the branch cuts but remains at a fixed distance from the singularities for any $\Lambda$ value. In this case, a careful investigation (not detailed here) leads to the conclusion that the $\Lambda\,{\to}\,0$ limit becomes interchangeable with the integrals, yielding finally
\begin{align}
&(*)\,{=}\,\sint{-3pt}{-12pt}0\infty\m d\beta\frac{f_1({\pm}1,\beta)}{\beta(\beta^2{+}1)}\slint{-2pt}{-2pt}\gamma\frac{\m da}{a{\mp}1}
\frac{a\,(\pm i{-}ia)^{\mp2i\eta}}{({-}ia)^{\mp2i\eta}(a{+}i)^2},
\end{align}
since there are no singularities above the line $\gamma$ on the $a$ plane, and we can close the path on the upper plane, so the $a$-integral yields zero by virtue of Cauchy's theorem.

In case of $\c A_{2\delta\pm}$, the branch cuts on the $b$ plane run from $B_r$ and $B_l$ in the $\m{Re}\,b\,{\le}\,0$ direction, and since the $b$-integrand decreases as $\sim\rec{b^2}$, and $\m{Im}(B'_1){\lessgtr}0$, and $|B'_1|{\ge}1$, we can tilt the $b$-path with a $\pm\frac\pi2$ angle (so that they run on the imaginary axis). Writing $b\,{=}\,{\pm}i\chi$, again one can convince oneself (with estimations similar to those used so far) that the limit and the integral is interchangeable in \Eq{e:A3dpm}. In this way we can perform the resulting $\chi$-integral:
\begin{align}
&\quad\c A_{2\delta\pm}=\pm i\sint{-2pt}{-9pt}{-1}1\m dy\frac{1{-}y}{1{+}y}\frac{f_2(1,y)}{1{+}y^2}\sint{-2pt}{-12pt}0\infty\m d\chi\frac{(\pm i\chi{+}1)^{\pm2i\eta-1}}{(\pm i\chi{-}1)^{\pm2i\eta+1}} = \nonumber\\
&\qquad = 
\frac{e^{2\pi\eta}{-}1}{\pm4i\eta}\sint{-2pt}{-9pt}{-1}1\m dy\frac{1{-}y}{1{+}y}\frac{f_2(1,y)}{1{+}y^2}.
\end{align}
This last integral has the $b{=}1$ sphere as its support; one can verify that it is indeed an integral over this sphere with respect to the usual spherical surface measure. This consideration, together with \Eq{e:A2dA2dpm} above, leads to the result stated in \Eq{e:A2dfinal} in the main text.

} 
\bibliographystyle{bst/sn-mathphys-num}
\bibliography{references}

\end{document}